\pdfoutput=1

\documentclass[11pt,twoside,a4paper,cmspaper,final,collab]{cms-tdr}
\def\svnVersion{474239P}\def\cmsCernNoTag{CERN-EP-2018-026}\def\cmsCernDate{\today}\def\cmsMessage{Published in the Journal of High Energy Physics as \href{http://dx.doi.org/10.1007/JHEP09(2018)007}{\doi{10.1007/JHEP09(2018)007.}}}

\begin{document}\cmsNoteHeader{HIG-17-020}

\hyphenation{had-ron-i-za-tion}
\hyphenation{cal-or-i-me-ter}
\hyphenation{de-vices}
\RCS$HeadURL: svn+ssh://svn.cern.ch/reps/tdr2/papers/HIG-17-020/trunk/HIG-17-020.tex $
\RCS$Id: HIG-17-020.tex 474237 2018-09-07 12:11:56Z rwolf $
\newlength\cmsFigWidth
\newcommand{\Wjets}{\ensuremath{\PW}{+}\text{jets}\xspace}
\newcommand{\ptem}{\ensuremath{\pt^{\Pe(\mu)}}\xspace}
\newcommand{\Ndata}{\ensuremath{N_{\text{data}}}}
\newcommand{\mTm}{\ensuremath{m_{\text{T}}^{\mu}}\xspace}
\newcommand{\mTem}{\ensuremath{m_{\text{T}}^{\Pe(\mu)}}\xspace}
\newcommand{\mTtot}{\ensuremath{m_{\text{T}}^{\text{tot}}}\xspace}
\newcommand{\Dzeta}{\ensuremath{D_{\zeta}}\xspace}
\newcommand{\Iabsem}{\ensuremath{I_{\text{abs}}^{\Pe(\mu)}}\xspace}
\newcommand{\Irelem}{\ensuremath{I_{\text{rel}}^{\Pe(\mu)}}\xspace}
\newcommand{\mvis}{\ensuremath{m_{\text{vis}}}\xspace}
\newcommand{\Pgth}{\Pgt_{\text{h}}\xspace}
\newcommand{\mA}{m_{\text{A}}}
\newcommand{\emu}{\Pe\Pgm}
\newcommand{\ee}{\Pe\Pe}
\newcommand{\mumu}{\Pgm\Pgm}
\newcommand{\mutau}{\Pgm\Pgth}
\newcommand{\etau}{\Pe\Pgth}
\newcommand{\tautau}{\Pgth\Pgth}
\newcommand{\mhmodp}{\ensuremath{m_{\text{h}}^{\text{mod+}}}}
\newcommand{\FF}{\ensuremath{F_{\text{F}}}}
\newlength\cmsTabSkip\setlength{\cmsTabSkip}{2ex}

\cmsNoteHeader{HIG-17-020}
\title{Search for additional neutral MSSM Higgs bosons in the $\Pgt\Pgt$ final state in proton-proton collisions at $\sqrt{s}=13\TeV$}

\date{\today}

\abstract{
  A search is presented for additional neutral Higgs bosons in the $\tau\tau$ final state in proton-proton collisions at the LHC. The search is performed in the context of the minimal supersymmetric extension of the standard model (MSSM), using the data collected with the CMS detector in 2016 at a center-of-mass energy of 13\TeV, corresponding to an integrated luminosity of 35.9\fbinv. To enhance the sensitivity to neutral MSSM Higgs bosons, the search includes production of the Higgs boson in association with b quarks. No significant deviation above the expected background is observed. Model-independent limits at 95\% confidence level (\CL) are set on the product of the branching fraction for the decay into $\tau$ leptons and the cross section for the production via gluon fusion or in association with b quarks. These limits range from 18\unit{pb} at 90\GeV to 3.5\unit{fb} at 3.2\TeV for gluon fusion and from 15\unit{pb} (at 90\GeV) to 2.5\unit{fb} (at 3.2\TeV) for production in association with b quarks, assuming a narrow width resonance. In the $\mhmodp$ scenario these limits translate into a 95\% \CL exclusion of $\tan\beta>6$ for neutral Higgs boson masses below 250\GeV, where $\tan\beta$ is the ratio of the vacuum expectation values of the neutral components of the two Higgs doublets. The 95\% \CL exclusion contour reaches 1.6\TeV for $\tan\beta=60$.
}

\hypersetup{
pdfauthor={CMS Collaboration},
pdftitle={Search for additional neutral MSSM Higgs bosons in the tau tau final state in proton-proton collisions at sqrt(s)=13 TeV},
pdfsubject={CMS},
pdfkeywords={CMS, physics, higgs, BSM, MSSM}}

\maketitle
\section{Introduction}

The discovery of a Higgs boson at the CERN LHC in 2012~\cite{Aad:2012tfa,Chatrchyan:2012xdj,Chatrchyan:2013lba}
has provided evidence that spontaneous symmetry breaking, as proposed by the Brout--Englert--Higgs
mechanism~\cite{Higgs:1964ia,Higgs:1964pj,Guralnik:1964eu,Higgs:1966ev,Kibble:1967sv,Englert:1964et},
may indeed be realized in nature. The determination of the properties of the new particle, based on
the complete LHC Run-1 data set~\cite{Aad:2015zhl,Khachatryan:2016vau}, has revealed
its consistency with the standard model (SM) Higgs boson, within the experimental accuracy. However
several questions remain, concerning, for example, the underlying mechanism responsible for the
symmetry breaking, or the exact form of the potential that breaks the symmetry.
To address these questions one of the main tasks of the LHC is the further exploration of the Higgs
sector. This includes the search for more complex structures, for example, in the form of more than
one Higgs doublet. Supersymmetry (SUSY)~\cite{Golfand:1971iw,Wess:1974tw} is an example of a beyond
the SM theory with a more complex Higgs sector. In the minimal supersymmetric standard model
(MSSM)~\cite{Fayet:1974pd,Fayet:1977yc} each particle of the SM is complemented by a SUSY partner,
which has the same properties apart from its spin. The Higgs sector of the MSSM consists of two
complex Higgs doublets, $\PH_{\text{u}}$ and $\PH_{\text{d}}$, to provide masses for up- and down-type
fermions. In the CP-conserving MSSM this leads to the prediction of five physical Higgs bosons:
two charged Higgs bosons $\PH^{\pm}$, two neutral scalar Higgs bosons $\Ph$ and $\PH$ (with masses
$m_{\Ph}<m_{\PH}$) and one neutral pseudoscalar Higgs boson A. At tree-level in the MSSM, the masses
of these five Higgs bosons and their mixing can be expressed in terms of the gauge boson masses and
two additional parameters, which can be chosen as the mass of the A, $\mA$, and the ratio of the
vacuum expectation values of the neutral components of the two Higgs doublets
\begin{linenomath}
  \begin{equation}
    \tan\beta = \frac{\langle \PH^{0}_{\text{u}}\rangle}{\langle \PH^{0}_{\text{d}}\rangle} =
    \frac{v_{\text{u}}}{v_{\text{d}}}.
  \end{equation}
\end{linenomath}
Dependencies on additional parameters of the SUSY breaking mechanism enter via higher-order
corrections in perturbation theory. In the exploration of the MSSM Higgs sector these parameters
are usually set to fixed values in the form of indicative benchmark scenarios~\cite{Carena:2013ytb}
to illustrate certain properties of the theory. For values of $m_{\text{A}}\gtrsim300\GeV$, which
seem to be favored by data~\cite{Aad:2015zhl,Khachatryan:2016vau,Aaboud:2017sjh,Khachatryan:2014wca},
the MSSM is close to the decoupling limit: the $\Ph$ usually takes the role of the observed SM-like
Higgs boson at 125\GeV and the $\PH$ and A are nearly degenerate in mass.

\begin{figure}[htbp]
  \centering
  \includegraphics[width=0.32\textwidth]{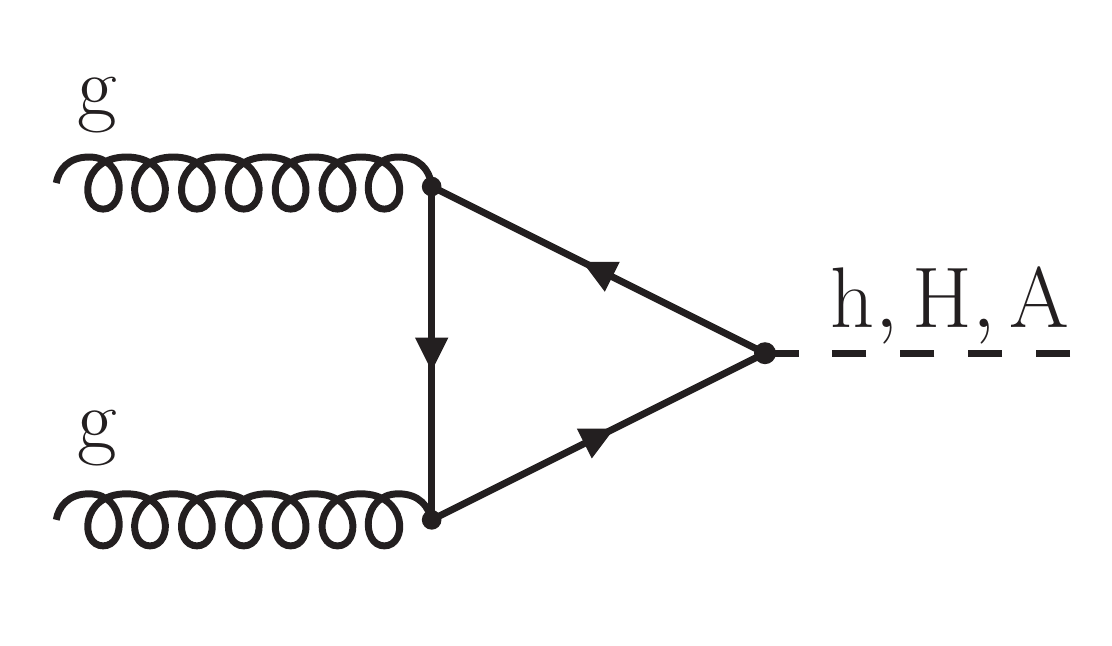}
  \includegraphics[width=0.32\textwidth]{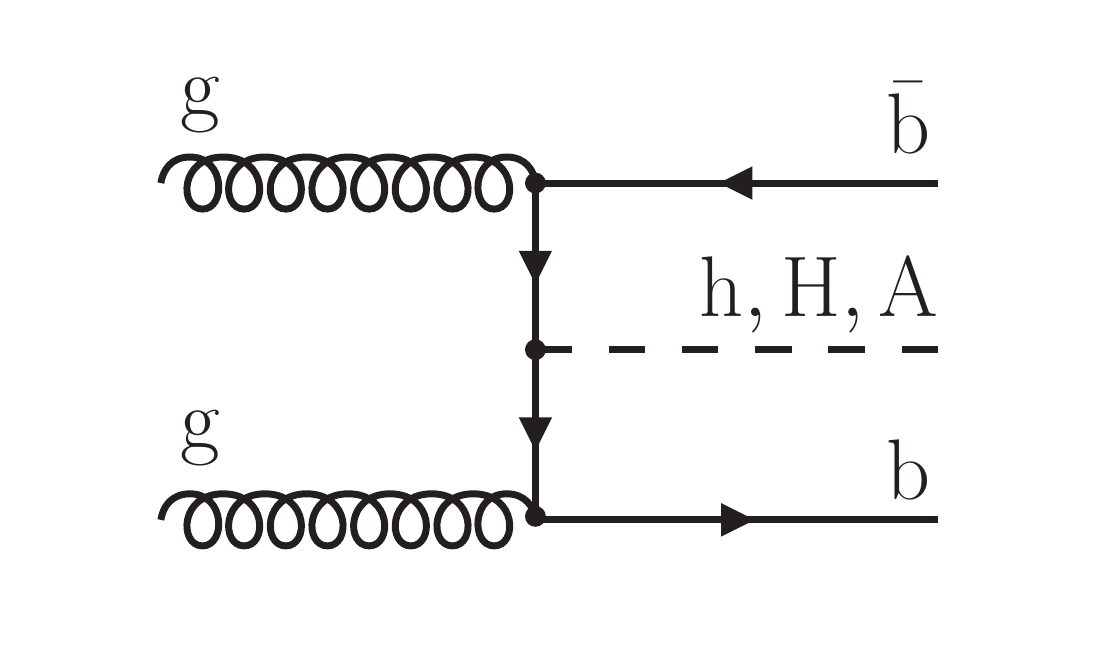}
  \includegraphics[width=0.32\textwidth]{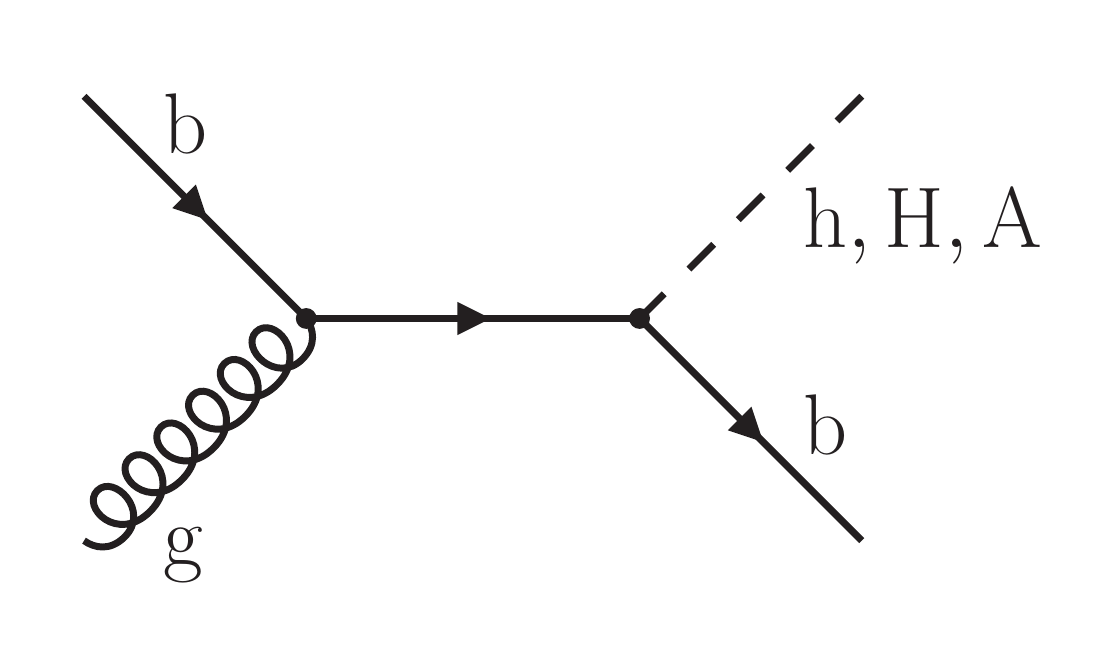}
  \caption {
    Diagrams for the production of neutral Higgs bosons (left) via gluon fusion and (middle and right)
    in association with b quarks. In supersymmetric extensions of the SM, the super-partners also
    contribute to the fermion loop, shown in the left panel. In the middle panel a pair of b quarks
    is produced from two gluons (the LO process in the four-flavor scheme). In the right panel the
    Higgs boson is radiated from a b quark in the proton (the LO process in the five-flavor scheme).
  }
  \label{fig:production-diagrams}
\end{figure}

At leading-order (LO), the coupling of the $\PH$ and the A to down-type fermions is enhanced by
$\tan\beta$ with respect to the expectation for an SM Higgs boson of the same mass, while the
coupling to vector bosons and up-type fermions is suppressed. The enhanced coupling to down-type
fermions makes searches for additional heavy neutral Higgs bosons that exploit final states containing
such fermions particularly interesting. It also has consequences for the production: firstly, the
production in association with b quarks dominates over the production via gluon fusion for large
values of $\tan\beta$. Secondly, in gluon fusion production the kinematic properties of the Higgs
boson change as a function of $\tan\beta$ due to the increasing contribution of b quarks in the
fermion loop. Diagrams for $\text{h}$, $\PH$, and A production at LO are shown in
Fig.~\ref{fig:production-diagrams}.

Searches for additional heavy neutral Higgs bosons in the context of the MSSM were carried out in
$\Pep\Pem$ collisions at LEP~\cite{Schael:2006cr} and in proton-antiproton collisions at the
Tevatron~\cite{Aaltonen:2009vf,Abazov:2010ci,Abazov:2011jh,Aaltonen:2011nh}. At the LHC such searches
have been carried out by the ATLAS and CMS Collaborations in the b quark~\cite{Chatrchyan:2013qga,
Khachatryan:2015tra,Sirunyan:2018taj}, dimuon~\cite{Aad:2012cfr,CMS:2015ooa}, and $\tau\tau$~\cite{Aad:2012cfr,
Aad:2014vgg,Aaboud:2016cre,Aaboud:2017sjh,Chatrchyan:2011nx,Chatrchyan:2012vp,Khachatryan:2014wca}
final states. The better experimental accessibility with respect to the b quark final state and the
larger mass, and therefore larger coupling, with respect to the muon give the $\tau\tau$ final state
a leading role in these searches.

In this paper the results of a search for additional heavy neutral Higgs bosons in the context of the
MSSM are presented. They are based on the 2016 $\Pp\Pp$ collision data set, taken at a center-of-mass
energy of 13\TeV, by the CMS experiment, and correspond to an integrated luminosity of $35.9\fbinv$.
The analysis is performed in four different $\tau\tau$ final states: $\emu$, $\etau$, $\mutau$, and
$\tautau$, where $\Pe$, $\mu$ and $\Pgth$ indicate $\tau$ lepton decays into electrons, muons and
hadrons respectively. For this analysis the most significant backgrounds are estimated from data, by
using new techniques with respect to previous publications by CMS. Upper limits are presented on the
product of the branching fraction for the decay into $\tau$ leptons and the cross section for the
production of a single narrow resonance via gluon fusion or in association with b quarks. In addition,
exclusion contours in the $\mA$-$\tan\beta$ plane in selected MSSM benchmark scenarios are provided.

In Sections~\ref{sec:detector} and~\ref{sec:event-reconstruction} the CMS detector and the event
reconstruction are described. Section~\ref{sec:event-selection} summarizes the event selection and
categorization. The event simulation and background estimation methods used for the analysis are
described in Section~\ref{sec:backgrounds}. The signal extraction is discussed in
Section~\ref{sec:signal-extraction}, followed by a discussion of the systematic uncertainties in
Section~\ref{sec:uncertainties}. Section~\ref{sec:results} contains the results of the analysis.
A summary is given in Section~\ref{sec:summary}.

\section{The CMS detector}
\label{sec:detector}

The central feature of the CMS apparatus is a superconducting solenoid of 6\unit{m} internal diameter,
providing a magnetic field of 3.8\unit{T}. Within the solenoid volume are a silicon pixel and strip
tracker, a lead tungstate crystal electromagnetic calorimeter (ECAL), and a brass and scintillator
hadron calorimeter (HCAL), each composed of a barrel and two endcap sections. Forward calorimeters
extend the pseudorapidity coverage provided by the barrel and endcap detectors. Muons are detected
in gas-ionization chambers embedded in the steel flux-return yoke outside the solenoid.

The silicon tracker measures charged particles within the pseudorapidity range $\abs{\eta} < 2.5$.
It consists of 1440 silicon pixel and $15\,148$ silicon strip detector modules. For nonisolated particles
with a transverse momentum of $1 < \pt < 10\GeV$ and $\abs{\eta} < 1.4$, the track resolutions
are typically 1.5\% in \pt and 25--90 (45--150)\unit{\mum} in the transverse (longitudinal) impact
parameter~\cite{TRK-11-001}. The electron momentum is estimated by combining the energy measurement
in the ECAL with the momentum measurement in the tracker. The momentum resolution for electrons with
$\pt \approx 45\GeV$ from $\PZ\to\Pe\Pe$ decays ranges from 1.7\% for nonshowering electrons in the
barrel region to 4.5\% for showering electrons in the endcaps~\cite{Khachatryan:2015hwa}. Muons are
measured in the pseudorapidity range $\abs{\eta} < 2.4$, with detection planes made using three
technologies: drift tubes, cathode strip chambers, and resistive-plate chambers. Matching muons to
tracks measured in the silicon tracker results in a relative \pt resolution for muons with $20 <\pt
<100\GeV$ of 1.3 to 2.0\% in the barrel and better than 6\% in the endcaps. The \pt resolution in
the barrel is better than 10\% for muons with \pt up to 1\TeV~\cite{Chatrchyan:2012xi}. In the
barrel section of the ECAL, an energy resolution of about 1\% is achieved for unconverted or
late-converting photons in the tens of GeV energy range. The remaining barrel photons have
a resolution of better than 2.5\% for $\abs{\eta} \leq 1.4$. In the endcaps, the resolution of
unconverted or late-converting photons is about 2.5\%, while the remaining endcap photons have a
resolution between 3 and 4\%~\cite{CMS:EGM-14-001}. When combining information from the entire
detector, the jet energy resolution amounts typically to 15\% at 10\GeV, 8\% at 100\GeV, and 4\% at
1\TeV, to be compared to about 40, 12, and 5\% obtained when the ECAL and HCAL calorimeters alone
are used.

Events of interest are selected using a two-tiered trigger system~\cite{Khachatryan:2016bia}. The
first level, composed of custom hardware processors, uses information from the calorimeters and
muon detectors to select events at a rate of around 100\unit{kHz} within a time interval of less than
4\unit{\mus}. The second level, known as the high-level trigger, consists of a farm of processors
running a version of the full event reconstruction software optimized for fast processing, and reduces
the event rate to around 1.0\unit{kHz} before data storage.

A more detailed description of the CMS detector, together with a definition of the coordinate system
used and the relevant kinematic variables, can be found in Ref.~\cite{Chatrchyan:2008zzk}.

\section{Event reconstruction}
\label{sec:event-reconstruction}

The reconstruction of the $\Pp\Pp$ collision products is based on the particle-flow (PF) algorithm as
described in Ref.~\cite{Sirunyan:2017ulk}, combining the available information from all CMS subdetectors
to reconstruct an unambiguous set of individual particle candidates. The particle candidates are
categorized into electrons, photons, muons, and charged and neutral hadrons. During the 2016 data
taking period the CMS experiment was operating with, on average, 23 inelastic $\Pp\Pp$ collisions per bunch
crossing. The fully recorded data of a bunch crossing defines an \emph{event} for further processing.
Collision vertices are obtained from reconstructed tracks using a deterministic annealing
algorithm~\cite{726788}. The reconstructed vertex with the largest value of summed physics-object
$\pt^{2}$ is taken to be the primary $\Pp\Pp$ interaction vertex. The physics objects for this purpose
are the jets, clustered using the jet finding algorithm~\cite{Cacciari:2008gp,Cacciari:2011ma}, as
described below, with the tracks assigned to the vertex as inputs, and the associated missing transverse
momentum, taken as the negative vector sum of the \pt of those jets. Any other collision vertices
in the event are associated with additional soft inelastic $\Pp\Pp$ collisions called \emph{pileup}.

Electrons are reconstructed by combining clusters of energy deposits in the ECAL with hits in the
tracker~\cite{Khachatryan:2015hwa}. To increase their purity, reconstructed electrons are required to
pass a multivariate electron identification discriminant, which combines information on track quality,
shower shape, and kinematic quantities. For this analysis working points with an efficiency between
80 and 90\% are used to identify electrons.
Muons in the event are reconstructed by performing a simultaneous track fit to hits in the tracker
and in the muon chambers~\cite{Chatrchyan:2012xi}. The presence of hits in the muon chambers already
leads to a strong suppression of particles misidentified as muons. Additional identification requirements
on the track fit quality and the compatibility of individual track segments with the fitted track can
reduce the misidentification rate further. For this analysis muon identification requirements with an
efficiency of ${\approx}$99\% are chosen. The contribution from backgrounds to the electron (muon)
selection is further reduced by requiring the corresponding lepton to be isolated from any hadronic
activity in the detector. This property is quantified by a relative isolation variable $\Irelem$,
which starts from the sum of the transverse momentum (energy) of all charged (neutral) particles,
$\Iabsem=\left(\sum p_{\text{T},i} + \sum E_{\text{T},i}\right)$ in a predefined cone of radius $\Delta
R = \sqrt{\smash[b]{\left(\Delta\eta\right)^{2}+\left(\Delta\phi\right)^{2}}}$ around the lepton
direction at the primary collision vertex, where $\Delta\eta$ and $\Delta\phi$ (measured in radians)
correspond to the angular distance of the particle to the lepton in the $\eta$ and $\phi$ directions.
The chosen cone size is $\Delta R<0.3 \,(0.4)$ for electrons (muons). The lepton itself is not included
in this calculation. To mitigate any distortions from pileup only those charged particles whose tracks
are associated with the primary collision vertex are taken into account. The presence of neutral
particles from pileup is estimated by summing the \pt of charged particles in the isolation cone
whose tracks have been associated to pileup vertices, and multiplying this quantity by a factor of
0.5 to account for the approximate ratio of neutral to charged hadron production. The value obtained
is subtracted from $\Iabsem$ and the result set to zero in case of negative values. Finally, $\Iabsem$
is divided by the \pt of the lepton to result in $\Irelem$.

For further characterization of the event all reconstructed PF objects are clustered into jets using
the anti-$\kt$ jet clustering algorithm as implemented in \textsc{fastjet}~\cite{Cacciari:2008gp,
Cacciari:2011ma} with a distance parameter of 0.4. To identify jets resulting from the hadronization
of b quarks a re-optimized version of the \emph{combined secondary vertex} b tagging algorithm, which
exploits information from the decay vertices of long-lived hadrons, and the impact parameters of
charged-particle tracks, in a combined discriminant, is used~\cite{BTV-16-002}. In the analysis a
working point corresponding to a b jet identification efficiency of ${\approx}$70\% and a
misidentification rate for light quarks and gluons of 1\% is chosen. Jets are also used as seeds for
the reconstruction of hadronic $\tau$ lepton decays. This is done by further exploiting the substructure
of the jets, using the \emph{hadrons-plus-strips} algorithm, as described in
Refs.~\cite{Khachatryan:2015dfa,CMS-PAS-TAU-16-002}. For the analysis the decay into three charged
hadrons and the decay into a single charged hadron accompanied by up to two neutral pions with $\pt>
2.5\GeV$ are used. The neutral pions are reconstructed as \emph{strips} with dynamic size from
reconstructed electrons and photons contained in the seeding jet, where the strip size varies as a
function of the \pt of the electron or photon candidate. The hadronic $\tau$ decay mode is then
obtained by combining the charged hadrons with the strips. Since they do not carry color charge,
high-$\pt$ $\tau$ leptons are expected to be isolated from any hadronic activity in the event as are
high-$\pt$ electrons and muons. Furthermore, in accordance with its finite lifetime the charged decay
products of the $\tau$ lepton are expected to be slightly displaced from the primary collision vertex.
To distinguish hadronic $\tau$ lepton decays from jets originating from the hadronization of quarks
or gluons a multivariate $\Pgth$ identification discriminant is used~\cite{Khachatryan:2015dfa}. It
combines information on the hadronic activity in the detector in the vicinity of the $\Pgth$ candidate
with the reconstructed lifetime information from the tracks of the charged decay products. Of the
predefined working points in Ref.~\cite{Khachatryan:2015dfa} this analysis makes use of the
\text{Tight}, \text{Medium}, and \text{VeryLoose} working points. These have efficiencies of 27\%
(\text{Tight}), 51\% (\text{Medium}), and 71\% (\text{VeryLoose}), for quark/gluon misidentification
rates of less then $4.4\times 10^{-4}$ (\text{Tight}), $3.3\times 10^{-3}$ (\text{Medium}), and $1.3
\times 10^{-2}$ (\text{VeryLoose}). Finally, requirements are imposed to reduce the misidentification
of electrons and muons as hadronic $\tau$ lepton decays. Also here predefined working points are used
to discriminate against electrons, with efficiencies ranging from 65\% (\text{Tight}) to 94\%
(\text{VeryLoose}) for electron misidentification rates between $6.2\times10^{-4}$ (\text{Tight}) and
$2.4\times 10^{-2}$ (\text{VeryLoose}). The misidentification rate of muons as hadronic $\tau$ lepton
decays is of $\mathcal{O}(10^{-3})$, for a $\Pgth$ identification efficiency of 99\%.

The missing transverse momentum vector \ptvecmiss, defined as the negative vector sum of the \pt of
all reconstructed PF objects, is also used to characterize the events. Its magnitude is referred to
as $\ptmiss$. It is used for the discrimination of backgrounds that are expected to contain neutrinos
with significantly more \pt than expected from the $\tau\tau$ final state, such as $\PW$ boson
production in association with jets ($\Wjets$). It is furthermore used for the calculation of the
final discriminating variable that is used for the statistical analysis, as detailed in
Section~\ref{sec:signal-extraction}.

\section{Event selection and categorization}
\label{sec:event-selection}

The four most sensitive final states of the $\tau\tau$ pair are exploited: $\emu$, $\etau$, $\mutau$,
and $\tautau$. The online selection for the $\etau$ ($\mutau$) final state is based on the presence
of at least one electron (muon) with  $\pt>25\,(22)\GeV$ and $\abs{\eta}<2.1$ at trigger level.
The online selection for the $\emu$ final state relies on a logical \emph{or} of two lower threshold
triggers that both require the presence of an electron and muon in the event with $\pt>23\GeV$ for
the higher-$\pt$ lepton and $\pt>12\,(8)\GeV$ for the lower-$\pt$ electron (muon). In the $\tautau$
final state, a trigger decision based on the presence of two hadronically decaying $\tau$ leptons with
$\pt>35\GeV$ and $\abs{\eta}<2.1$ is used.

\begin{table}[h]
  \topcaption{
    Kinematic selection of the $\tau$ lepton decay products in the $\emu$, $\etau$, $\mutau$, and
    $\tautau$ final states. The expression ``First (Second) object'' refers to the final state label
    used in the first column.
  }
  \label{tab:lepton-selection}
  \centering
  \begin{tabular}{lr@{$>$}l@{, }r@{$<$}l@{\hspace{0.6cm}}r@{$>$}l@{, }r@{$<$}l}
    Final state & \multicolumn{4}{c}{First object} & \multicolumn{4}{c}{Second object} \\
    \hline
    $\emu^{\dagger}$       & $\pt^{\Pe\hspace{0.05cm}}$ & 13\GeV & $\abs{\eta^{\Pe\hspace{0.05cm}}}$ & 2.5
    & $\pt^{\mu\hspace{0.125cm}}$ & 10\GeV & $\abs{\eta^{\mu\hspace{0.125cm}}}$ & 2.4 \\
    $\etau$       & $\pt^{\Pe\hspace{0.05cm}}$ & 26\GeV & $\abs{\eta^{\Pe\hspace{0.05cm}}}$ & 2.1
    & $\pt^{\Pgth}$ & 30\GeV & $\abs{\eta^{\Pgth}}$ & 2.3 \\
    $\mutau$      & $\pt^{\mu}$ & 23\GeV & $\abs{\eta^{\mu}}$ & 2.1 & $\pt^{\Pgth}$ & 30\GeV
    & $\abs{\eta^{\Pgth}}$ & 2.3 \\
    $\tautau$ & \multicolumn{8}{c}{$\pt^{\Pgth}>40\GeV$, $\abs{\eta^{\Pgth}}<2.1$} \\
    \hline
    \multicolumn{9}{l}{\small $^{\dagger}$ For events passing only one trigger an additional requirement
    of $\pt> 24\GeV$ is} \\
    \multicolumn{9}{l}{\small $\hphantom{^{\dagger}}$ applied on the higher-$\pt$ lepton candidate as
    explained in the text.}
  \end{tabular}
\end{table}

Requirements on the \pt and $\eta$ of the reconstructed $\tau$ lepton decay products are applied
in the offline analysis as given in Table~\ref{tab:lepton-selection}. In the $\emu$ final state an
electron with $\pt>13\GeV$ and $\abs{\eta}<2.5$ and a muon with $\pt>10\GeV$ and $\abs{\eta}
<2.4$ are required. If the event passed only one trigger the lepton identified with the higher-$\pt$
trigger object is required to have a $\pt>24\GeV$. This guarantees a trigger acceptance well above
the turn-on of at least one of the triggers used. Both leptons are required to pass identification
criteria as described in Section~\ref{sec:event-reconstruction} and to be isolated according to
$\Irelem<0.15\,(0.2)$. Events with additional electrons or muons fulfilling looser selection
requirements than these are rejected.

In the $\etau$ ($\mutau$) final state an electron (muon) with $\pt>26\,(23)\GeV$ and $\abs{
\eta}<2.1$ and a $\Pgth$ candidate with $\pt>30\GeV$ and $\abs{\eta}<2.3$ are required. The electron
(muon) and the $\Pgth$ candidate should fulfill the identification requirements as described in
Section~\ref{sec:event-reconstruction}. The $\Pgth$ candidate should pass the \text{Tight} working
point of the $\Pgth$ identification discriminant, the \text{Tight} (\text{VeryLoose}) working point
of the discriminant to suppress electrons and the \text{Loose} (\text{Tight}) working point of the
discriminant to suppress muons in the $\etau$ ($\mutau$) case. In addition, the electron (muon)
should be isolated according to $\Irelem<0.1\,(0.15)$. Events with additional electrons or muons
fulfilling looser selection requirements are rejected.

In the $\tautau$ final state two $\Pgth$ candidates with $\pt>40\GeV$ and $\abs{\eta}<2.1$ are
required. Both must pass the \text{Medium} working point of the $\Pgth$ identification discriminant,
the VeryLoose working point of the discriminant against electrons and the \text{Loose} working point
of the discriminant against muons. Events with additional electrons or muons fulfilling looser
requirements on identification, isolation and \pt than described for the $\etau$ or $\mutau$ final
state above are rejected.

In all cases the decay products of the two $\tau$ leptons are required to be of opposite charge,
separated by more than 0.5 in $\Delta R$ and associated to the primary collision vertex within a
distance of $0.045$\unit{cm} in the transverse plane for electrons and muons and $0.2$\unit{cm} along
the beam axis for all final-state particles. The vetoing of additional electrons or muons helps with
the suppression of backgrounds and ensures that no event will be categorized according to more than
one $\tau\tau$ final state. At most 0.8\% of the selected events contain more $\Pgth$ candidates than
required for the corresponding final state. In this case, the $\tau\tau$ pair with the most isolated
final state products is chosen.

To increase the sensitivity of the analysis all selected events are further categorized: events with
at least one jet with $\pt>20\GeV$ and $\abs{\eta}<2.4$ that passes the b tagging requirement as described
in Section~\ref{sec:event-reconstruction} are combined into a global \text{b-tag} category. This
category is designed to target the production of the Higgs boson in association with b quarks. All
other events are added to a global \text{no b-tag} category.

In the $\emu$ final state each event category is further split into three subcategories based on the
quantity $\Dzeta$, introduced for the first time in Ref.~\cite{Abulencia:2005kq}, defined as
\begin{linenomath}
  \begin{equation}
    \label{eqn:Dzeta}
    \begin{split}
      \Dzeta &= p_{\zeta}^\text{miss} - 0.85\,p_{\zeta}^\text{vis} ; \qquad
      p_{\zeta}^\text{miss} = \ptvecmiss\cdot\hat{\zeta} ; \qquad
      p_{\zeta}^\text{vis} = \left(\vec{p}_{\text{T}}^{\kern1pt\Pe} + \vec{p}_{\text{T}}^{\kern1pt\mu}\right)
      \cdot\hat{\zeta}, \\
    \end{split}
  \end{equation}
\end{linenomath}
where $\vec{p}_{\text{T}}^{\,\Pe(\mu)}$ corresponds to the transverse momentum vector of the
electron (muon) and $\hat{\zeta}$ to the bisectional direction between the electron and the muon in
the transverse plane. The variables $p_{\zeta}^\text{miss}$ and $p_{\zeta}^\text{vis}$ in
Eq.~(\ref{eqn:Dzeta}) can take positive or negative values. The linear combination of $p_{\zeta
}^\text{miss}$ and $p_{\zeta}^\text{vis}$ has been chosen to optimize the sensitivity of the
analysis in the $\emu$ final state. The variable $\Dzeta$ is especially suited to suppress $\Wjets$
and $\ttbar$ events, where the reconstructed lepton candidates and the direction of $\ptvecmiss$ are
distributed more isotropically in the detector than for genuine $\tau\tau$ signal events. The
categories are defined as low-$\Dzeta$ ($-50<\Dzeta\leq -10\GeV$), medium-$
\Dzeta$ ($-10<\Dzeta\leq 30\GeV$) and high-$\Dzeta$ ($\Dzeta>30\GeV$). In this
way categories with different fractions of signal and $\ttbar$ events can be exploited for the statistical
analysis. The expected signal, for all masses tested, is mostly located in the medium-$\Dzeta$
subcategory.

In the $\etau$ ($\mutau$) final state each global event category is further split into two subcategories
based on the transverse mass of the electron or muon and $\ptmiss$
\begin{linenomath}
  \begin{equation}
    \label{eqn:mt_definition}
    \mTem = \sqrt{2\,\ptem\,\ptmiss\left(1 - \cos\Delta\phi\right)}.
  \end{equation}
\end{linenomath}
This transverse mass is used to discriminate between the signal and the backgrounds from $\Wjets$ and
$\ttbar$ events. In Eq.~(\ref{eqn:mt_definition}) $\ptem$ refers to the \pt of the electron (muon)
and $\Delta\phi$ to the difference in the azimuthal angle between the electron (muon) and $\ptvecmiss$.
The categories are defined as \text{tight-}$\mT$ ($\mTem<40\GeV$) and \text{loose-}$\mT$ ($40<\mTem<
70\GeV$). The bulk of the signal events, particularly for the low-mass hypotheses, lie in the tight-$
\mT$ subcategory. The \text{loose-}$\mT$ category has been added to increase the signal acceptance
for mass hypotheses of $m_{\text{A},\PH}>700\GeV$.

\begin{figure}[htbp]
  \centering
  \includegraphics[width=0.48\textwidth]{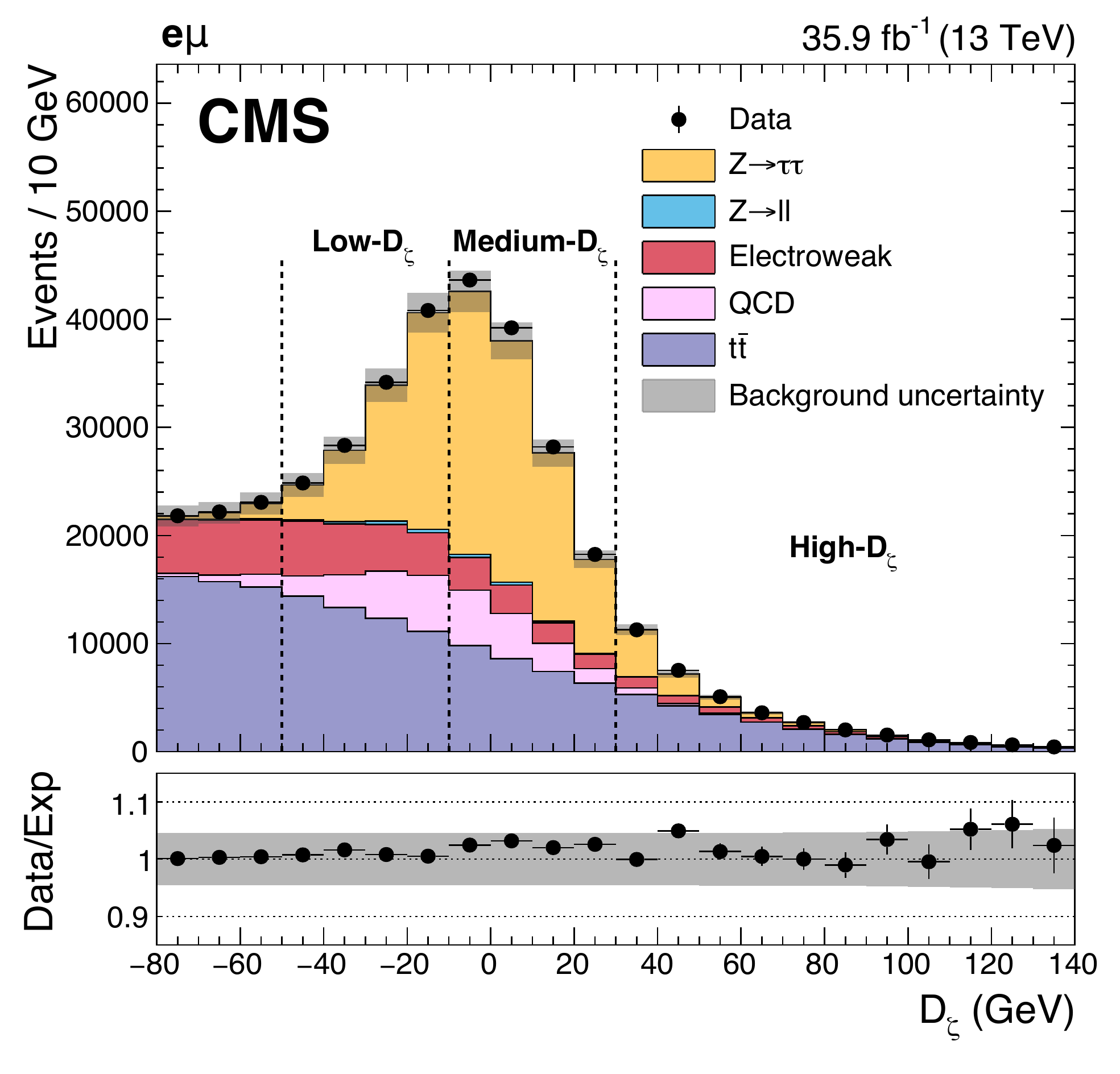}
  \includegraphics[width=0.48\textwidth]{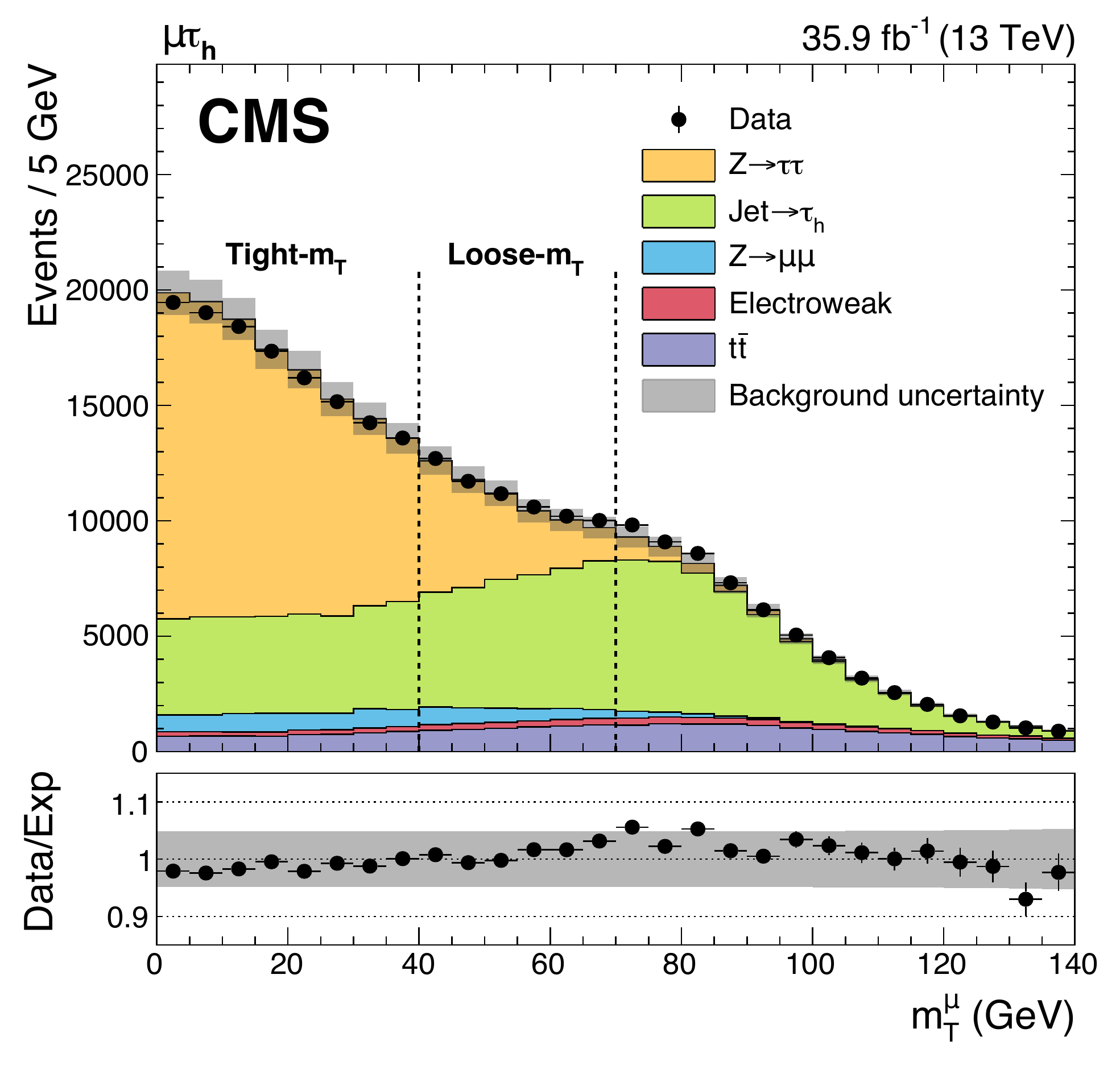}
  \caption {
    Observed and expected distributions of (left) $\Dzeta$ in the $\emu$ final state and (right)
    $\mTm$ in the $\mutau$ final state. The dashed vertical lines indicate the definition of the
    subcategories in each final state. The label ``$\text{jet}\to\Pgth$'' indicates events with jets
    misidentified as hadronic $\Pgt$ lepton decays, \eg $\Wjets$ events, which are estimated from
    data as described in Section~\ref{subsec:Data-backgrounds}. A detailed description of the
    composition of the expected background is given in Section~\ref{sec:signal-extraction}. The
    distributions are shown before any event categorization and prior to the fit used for the signal
    extraction. For these figures no uncertainties that affect the shape of the distributions have
    been included in the uncertainty model.
  }
  \label{fig:sub-categories}
\end{figure}

\begin{figure}[htbp]
  \centering
  \includegraphics[width=1.\textwidth]{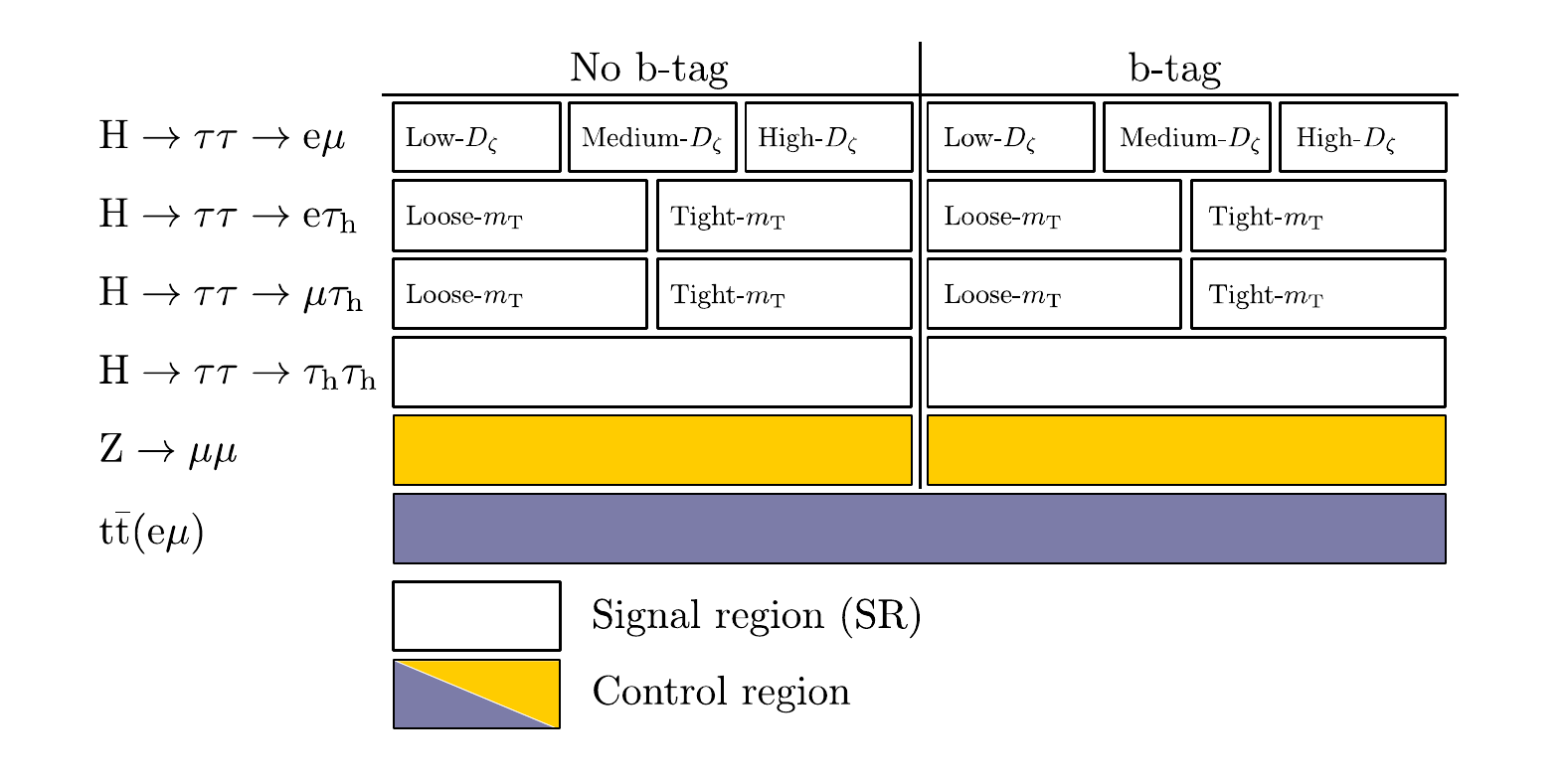}
  \caption {
    Overview of all event subcategories that enter the statistical analysis. Sixteen signal categories
    are complemented by three background control regions in the main analysis as described in
    Section~\ref{sec:backgrounds}.
  }
  \label{fig:categories}
\end{figure}

In combination this leads to 16 event categories entering the statistical analysis, complemented by
three background control regions, as discussed in Section~\ref{sec:backgrounds}. In
Fig.~\ref{fig:sub-categories}, the $\Dzeta$ and $\mTm$ distributions are shown in the $\emu$ and
$\mutau$ final states respectively, before splitting the events into categories, indicating the
corresponding subcategoriza\-tion. A discussion of the composition of the expected background
contributions is given in Section~\ref{sec:signal-extraction}. A graphical representation of the
complete event categorization is given in Fig.~\ref{fig:categories}.

\section{Event simulation and background estimation}
\label{sec:backgrounds}

A list of all SM backgrounds that contribute to the event selection described in
Section~\ref{sec:event-selection} is given in Table~\ref{tab:bg-processes}. The most obvious background
originates from $\PZ$ boson production in the $\tau\tau$ final state ($\PZ\to\tau\tau$). Since the
analysis is not sensitive to the CP-eigenvalue or spin of the Higgs boson, the signal can be distinguished
from this background only by the difference in mass of the associated bosons.  The same is true for
$\PZ\to\ell\ell$ events, where $\ell$ refers to an electron or muon, if one of the leptons is
misidentified as a hadronic $\tau$ lepton decay. Similar arguments hold for $\ttbar$ production (a
dominant background especially in the $\emu$ final state), the production of single $\cPqt$ quarks and
vector boson pair production ($\PW\PW$, $\PW\PZ$, and $\PZ\PZ$). Common to all these backgrounds in
the $\etau$, $\mutau$, and $\tautau$ final states is that they can be misinterpreted as signal events
in two ways: firstly, if the final state contains one or more genuine $\tau$ leptons or if an electron
or muon in the final state is misinterpreted as a hadronic $\tau$ lepton decay, and secondly, if one
or more jets are misinterpreted as hadronic $\tau$ lepton decays. In Table~\ref{tab:bg-processes} the
former is labeled as ``$\tau/\ell\to\Pgth$'', whilst the latter is labeled as ``$\text{jet}\to\Pgth$''.
Typical misidentification probabilities are given in Section~\ref{sec:event-reconstruction}. Backgrounds
due to $\Wjets$ or SM events comprised uniquely of jets produced through the strong interaction,
referred to as quantum chromodynamics (QCD) multijet production, predominantly contribute to the
event selection via the misidentification of jets as hadronic $\tau$ lepton decays. The level to
which each of these processes contributes to the event selection depends on the final state.

\begin{table}[h]
  \topcaption{
    Background processes contributing to the event selection, as given in Section~\ref{sec:event-selection}.
    The first row corresponds to the SM Higgs boson in the $\tau\tau$ final state, which is also taken
    into account in the statistical analysis. The further splitting of the processes in the second
    column refers only to final states that contain a $\Pgth$ candidate. The label ``MC'' implies
    that the process is taken from simulation; the label ``\FF'' implies that the process is determined
    from data using the fake factor method, as described in Section~\ref{subsec:Data-backgrounds}. The
    label ``CR'' implies that both the shape and normalization of QCD multijet events are estimated
    from control regions in data. The symbol $\ell$ corresponds to an electron or muon.
  }
  \label{tab:bg-processes}
  \centering
  \begin{tabular}{lr@{$\;\to\;$}lcccc}
    Background process & \multicolumn{2}{c}{Misidentification} & $\emu$ & $\etau$ & $\mutau$ & $\tautau$ \\
    \hline
    \multirow{2}{*}{$\PH\to\tau\tau$ (SM)} & \multicolumn{2}{l}{ } & \multirow{2}{*}{MC$\hphantom{^{\dagger}}$} &
    \multirow{2}{*}{MC$\hphantom{^{\dagger}}$} & \multirow{2}{*}{MC$\hphantom{^{\dagger}}$} &
    \multirow{2}{*}{MC$\hphantom{^{\dagger}}$} \\[\cmsTabSkip]
    \multirow{2}{*}{$\PZ\to\tau\tau$} & \multicolumn{2}{l}{ } & \multirow{2}{*}{MC$^{\dagger}$} & \multirow{2}{*}{MC$^{\dagger}$} &
    \multirow{2}{*}{MC$^{\dagger}$} & \multirow{2}{*}{MC$^{\dagger}$} \\[\cmsTabSkip]
    \multirow{2}{*}{$\PZ\to\ell\ell$} & $\ell$&$\Pgth$ & \multirow{2}{*}{MC$\hphantom{^{\dagger}}$} & MC$\hphantom{^{\dagger}}$
    & MC$\hphantom{^{\dagger}}$ & MC$\hphantom{^{\dagger}}$ \\
    & $\text{Jet}$&$\Pgth$    &  & \FF$\hphantom{^{\dagger}}$ & \FF$\hphantom{^{\dagger}}$ & \FF$\hphantom{^{\dagger}}$ \\[\cmsTabSkip]
    \multirow{2}{*}{Diboson+single $\cPqt$}     & $\tau/\ell$&$\Pgth$ & \multirow{2}{*}{MC$\hphantom{^{\dagger}}$} & MC$\hphantom{^{\dagger}}$
    & MC$\hphantom{^{\dagger}}$ & MC$\hphantom{^{\dagger}}$ \\
    & $\text{Jet}$&$\Pgth$   &  & \FF$\hphantom{^{\dagger}}$ & \FF$\hphantom{^{\dagger}}$ & \FF$\hphantom{^{\dagger}}$ \\[\cmsTabSkip]
    \multirow{2}{*}{$\ttbar$}               & $\hspace{0.45cm}\tau/\ell$&$\Pgth$ & \multirow{2}{*}{MC$^{\dagger}$} & MC$^{\dagger}$ & MC$^{\dagger}$
    & MC$^{\dagger}$ \\
    & $\text{Jet}$&$\Pgth$    &  & \FF$\hphantom{^{\dagger}}$ & \FF$\hphantom{^{\dagger}}$ & \FF$\hphantom{^{\dagger}}$ \\[\cmsTabSkip]
    \multirow{2}{*}{$\Wjets$} & \multicolumn{2}{c}{\multirow{2}{*}{$\hspace{0cm}\text{Jet}\to\Pgth$}} & \multirow{2}{*}{MC$\hphantom{^{\dagger}}$}
    & \multirow{2}{*}{\FF$\hphantom{^{\dagger}}$} & \multirow{2}{*}{\FF$\hphantom{^{\dagger}}$} & \multirow{2}{*}{\FF$\hphantom{^{\dagger}}$} \\[\cmsTabSkip]
    \multirow{2}{*}{QCD multijet production} & \multicolumn{2}{c}{\multirow{2}{*}{$\hspace{0cm}\text{Jet}\to\Pgth$}} & \multirow{2}{*}{CR$\hphantom{^{\dagger}}$}
    & \multirow{2}{*}{\FF$\hphantom{^{\dagger}}$} & \multirow{2}{*}{\FF$\hphantom{^{\dagger}}$} & \multirow{2}{*}{\FF$\hphantom{^{\dagger}}$} \\[\cmsTabSkip]
    \hline
    \multicolumn{6}{l}{\small $^{\dagger}$ Normalization from control region in data.} \\
  \end{tabular}
\end{table}

\subsection{Event simulation}
\label{subsec:simulation}

Drell--Yan events in the dielectron, dimuon, and $\tau\tau$ final states, and $\Wjets$ events are
generated at LO precision in the coupling strength $\alpha_{s}$~\cite{Alwall:2011uj}, using the
\MGvATNLO 2.2.2 event generator~\cite{Alwall:2014hca}. To increase the number of
simulated events in regions of high signal purity supplementary samples are generated with up to four
outgoing partons in the hard interaction. For diboson production \MGvATNLO is used
at next-to-leading order (NLO) precision. For $\ttbar$ and single $\cPqt$ quark production samples are
generated at NLO precision using \POWHEG 2.0~\cite{Nason:2004rx,Frixione:2007vw,Alioli:2008tz,
Alioli:2010xd,Alioli:2010xa,Bagnaschi:2011tu}. For the interpretation of the results the expected
contribution of the SM Higgs boson is taken into account; this process is simulated using \POWHEG
separately for the production via gluon fusion, vector boson fusion (VBF), or in association with
a $\PZ$ ($\PZ\PH$) or $\PW$ ($\PW\PH$) boson. When compared to data and not modified by a control
measurement in data, Drell--Yan, $\Wjets$, $\ttbar$, and single $\cPqt$ quark events in the t$\PW$-channel
are normalized to their cross sections at next-to-next-to-leading order (NNLO)
precision~\cite{Melnikov:2006kv,Czakon:2011xx,Kidonakis:2013zqa}. Single $\cPqt$ quark production in the
$t$-channel and diboson events are normalized to their cross sections at NLO
precision~\cite{Kidonakis:2013zqa,Campbell:2011bn}.
The gluon fusion signal process is simulated at LO precision using \PYTHIA 8.212~\cite{Sjostrand:2014zea}.
For the statistical analysis the Higgs boson $\pt$ distribution is weighted to NLO precision using
\POWHEG. To account for the multiscale nature of the process in the NLO plus parton shower \POWHEG
prediction, the $\pt$ spectra corresponding to the contributions from the t quark alone, the b quark
alone and the tb-interference are each calculated separately, using a \POWHEG damping factor set to
the individual scales as discussed in Refs.~\cite{Bagnaschi:2015bop,Bagnaschi:2015qta,Harlander:2014uea}.
For the model-independent limits the individual distributions are combined according to their
contribution to the total cross section as expected for a CP-even Higgs boson with given mass in the
SM. In the model-dependent interpretation in the MSSM, where the contributions of the individual
distributions also depend on the model parameters, these contributions are obtained using \POWHEG in
the two Higgs doublet mode. Each distribution is scaled, depending on the model parameters, using the
effective Yukawa couplings as predicted by the corresponding benchmark model, before all distributions
are combined into one single prediction. In this context also the $\tan\beta$ enhanced SUSY corrections
to the b quark coupling are taken into account via the corresponding effective Yukawa coupling where
appropriate. Other SUSY contributions have been checked to be less than a few percent and are
neglected. The associated production with b quarks is simulated at NLO precision using
\MGvATNLO~\cite{Wiesemann:2014ioa}.

For the generation of all signal and background processes the NNPDF3.0 parton distribution functions
(PDFs) are used, as described in Ref.~\cite{Ball:2014uwa}. The description of the underlying event is
parametrized according to the CUETP8M1 tune~\cite{Khachatryan:2015pea}. Hadronic showering and
hadronization, as well as the $\tau$ lepton decays, are modeled using \PYTHIA. For all simulated
events the effect of the observed pileup is taken into account. For this purpose additional inclusive
inelastic $\Pp\Pp$ collisions are generated with \PYTHIA and added to all simulated events according
to the expected pileup profile. All events generated are passed through a \GEANTfour-based~\cite{Agostinelli:2002hh}
simulation of the CMS detector and reconstructed using the same version of the CMS event reconstruction
software as used for the data. The observed event yields in each event category and the composition
of the expected background contributions to the selected events are given in Table~\ref{tab:bg-composition}.

\begin{sidewaystable}
  \centering
  \topcaption{
    Observed number of selected events ($\Ndata$) and the relative contribution of the
    expected backgrounds in all event categories in the $\emu$, $\etau$, $\mutau$, and $\tautau$ final
    states. The relative contribution of the expected backgrounds is given in \%, including the
    contribution of an SM Higgs boson with a mass of 125\GeV, and prior to the fit used for the
    signal extraction. In all but the $\emu$ final state, processes in which a jet is misidentified
    as a hadronic $\tau$ lepton decay are subsumed into a common $\text{jet}\to\Pgth$
    background class which is estimated from data.
  }
  \label{tab:bg-composition}
  \vspace{0.5cm}
  \begin{tabular}{lllrccccccc}
    \multicolumn{3}{c}{Category} & $\Ndata$ & $\PH\to\tau\tau$ (SM) & $\PZ\to\tau\tau$ & $\PZ\to\ell\ell\hphantom{^{dagger}}$
    & Diboson+single $\cPqt$$\hphantom{^{\dagger}}$ & $\ttbar\hphantom{^{dagger}}$ & $\Wjets$ & QCD \\
    \hline
    \multirow{6}{*}{$\emu$} & \multirow{3}{*}{\text{No b-tag}} & \text{Low}-$\Dzeta$ & 85816 & $\hphantom{0}0.24$
    & 49.55 & $\hphantom{0}1.06$ & 11.50 & 15.73 & 6.03 & 15.88 \\
    & & \text{Medium}-$\Dzeta$ & 102143 & $\hphantom{0}0.37$
    & 69.90 & $\hphantom{0}0.75$ & $\hphantom{0}4.97$ & $\hphantom{0}8.22$ & 1.94 & 13.85 \\
    & & \text{High}-$\Dzeta$ & 18364 & $\hphantom{0}0.50$
    & 45.66 & $\hphantom{0}0.38$ & 11.95 & 32.59 & 1.94 & $\hphantom{0}6.99$ \\
    & \multirow{3}{*}{$\hspace{0.63cm}$\text{b-tag}} & \text{Low}-$\Dzeta$ & 42870 & $\hphantom{0}0.01$
    & $\hphantom{0}2.08$ & $\hphantom{0}0.08$ & $\hphantom{0}6.88$ & 89.19 & 0.27 & $\hphantom{0}1.49$ \\
    & & \text{Medium}-$\Dzeta$ & 27803 & $\hphantom{0}0.03$
    & $\hphantom{0}4.37$ & $\hphantom{0}0.12$ & $\hphantom{0}6.57$ & $86.63$ & $0.14$ & $\hphantom{0}2.14$ \\
    & & \text{High}-$\Dzeta$ & 20208 & $\hphantom{0}0.03$
    & $\hphantom{0}1.72$ & $\hphantom{0}0.01$ & $\hphantom{0}7.00$ & 91.09 & 0.16 & $>0.01$ \\
    \multicolumn{11}{c}{ } \\
    \multicolumn{3}{c}{Category} & $\Ndata$ & $\PH\to\tau\tau$ (SM) & $\PZ\to\tau\tau$ & $\PZ\to\ell\ell^{\dagger}$
    & Diboson+single $\cPqt$$^{\dagger}$ & $\ttbar^{\dagger}$ & \multicolumn{2}{c}{$\text{jet}\to\Pgth$} \\
    \hline
    \multirow{4}{*}{$\etau$} & \multirow{2}{*}{\text{No b-tag}} & \text{Tight}-$\mT$ & 49120 & $\hphantom{0}0.66$
    & 55.82 & 12.72 & $\hphantom{0}0.95$ & $\hphantom{0}1.24$ & \multicolumn{2}{c}{28.61} \\
    & & \text{Loose}-$\mT$ & 29108 & $\hphantom{0}0.35$
    & 28.41 & 11.71 & $\hphantom{0}1.70$ & $\hphantom{0}2.32$ & \multicolumn{2}{c}{55.50} \\
    & \multirow{2}{*}{$\hspace{0.63cm}$\text{b-tag}} & \text{Tight}-$\mT$ & 3886 & $\hphantom{0}0.28$
    & 17.21 & $\hphantom{0}2.08$ & $\hphantom{0}4.97$ & 50.45 & \multicolumn{2}{c}{25.01} \\
    & & \text{Loose}-$\mT$ & 3657 & $\hphantom{0}0.10$ & $\hphantom{0}5.58$ & $\hphantom{0}1.65$
    & $\hphantom{0}5.65$ & 57.51 & \multicolumn{2}{c}{29.52} \\
    \multirow{4}{*}{$\mutau$} & \multirow{2}{*}{\text{No b-tag}} & \text{Tight}-$\mT$ & 124707 & $\hphantom{0}0.64$
    & 67.72 & $\hphantom{0}4.64$ & $\hphantom{0}0.88$ & $\hphantom{0}1.11$& \multicolumn{2}{c}{25.01} \\
    & & \text{Loose}-$\mT$ & 58534 & $\hphantom{0}0.38$
    & 34.62 & $\hphantom{0}5.77$ & $\hphantom{0}1.96$ & $\hphantom{0}2.57$ & \multicolumn{2}{c}{54.70} \\
    & \multirow{2}{*}{$\hspace{0.63cm}$\text{b-tag}} & \text{Tight}-$\mT$ & 8778 & $\hphantom{0}0.28$
    & 19.33 & $\hphantom{0}0.98$ & $\hphantom{0}4.80$ & $50.11$ & \multicolumn{2}{c}{24.50} \\
    & & \text{Loose}-$\mT$ & 7887 & $\hphantom{0}0.08$ & $\hphantom{0}5.86$ & $\hphantom{0}0.59$
    & $\hphantom{0}5.55$ & 59.65 & \multicolumn{2}{c}{28.28} \\
    \multirow{2}{*}{$\tautau$} & \text{No b-tag} & & 105545 & $\hphantom{0}0.45$ & 14.94 & $\hphantom{0}2.08$
    & $\hphantom{0}0.28$ & $\hphantom{0}0.23$ & \multicolumn{2}{c}{82.02} \\
    & $\hspace{0.63cm}$\text{b-tag} & & 3416 & $\hphantom{0}0.52$ & 20.80 & $\hphantom{0}1.03$
    & $\hphantom{0}2.65$ & 18.75 & \multicolumn{2}{c}{56.25} \\
    \hline
    \multicolumn{11}{l}{\small $^{\dagger}$ Excluding events with jets misidentified as hadronic $\Pgt$ lepton decays.} \\
  \end{tabular}
\end{sidewaystable}

\subsection{Backgrounds estimated from data}
\label{subsec:Data-backgrounds}

A large fraction of the backgrounds outlined for the $\etau$, $\mutau$, and $\tautau$ final states
in Table~\ref{tab:bg-processes} can be attributed to jets misidentified as hadronic $\tau$ lepton
decays. For the signal extraction, the shape and normalization of these backgrounds are estimated from
control regions in data, using the ``fake factor'' method, as described in Ref.~\cite{Sirunyan:2018qio}.
In this approach the number of events for a certain background $i$ due to $\text{jet}\to\Pgth$
misidentification is estimated from a region that only differs from the signal region (SR) by
modifying the $\Pgth$ identification requirement. This region is referred to as the application region
(AR). For this purpose the $\Pgth$ identification is required to fulfill the \text{VeryLoose} but not
the \text{Tight} (\text{Medium}) working point of the discriminant in the $\etau$/$\mutau$ ($\tautau$)
final state. This region is primarily populated by events with jets misidentified as hadronic $\tau$
lepton decays, with typical fractions of genuine $\tau$ lepton decays at the level of a few percent
or below. To arrive at an estimate for the number of events from background $i$ due to $\text{jet}\to
\Pgth$ misidentification in the SR the number of events in the AR is then multiplied by the ratio
\begin{linenomath}
  \begin{equation}
    \FF^{i} = \frac{N_{\text{pass}}}{N_{\text{fail}}},
    \label{eq:fake-factor}
  \end{equation}
\end{linenomath}
where $N_{\text{pass}}$ corresponds to the number of events that fulfill the \text{Tight/Medium}
working point and $N_{\text{fail}}$ to the number of events that fulfill the \text{VeryLoose} but
not the \text{Tight/Medium} working point of the $\Pgth$ identification discriminant. The number of
events appearing in Eq.~(\ref{eq:fake-factor}) is obtained from a dedicated determination region
(DR$_{i}$), which is orthogonal to the AR and SR, and dominated by background $i$. The contributions
from backgrounds other than $i$ are estimated from simulation and subtracted from the numerator and
denominator of Eq.~(\ref{eq:fake-factor}). For this purpose all corrections as described in
Section~\ref{subsec:MC-backgrounds} are applied to the simulation. The $\FF^{i}$ can be different
for different processes, for example, if the misidentified jet predominantly originates from a heavy
flavor quark, a light flavor quark or gluon fragmentation.

\begin{figure}[htbp]
  \centering
  \setlength{\unitlength}{\textwidth}
  \includegraphics[width=.7\textwidth]{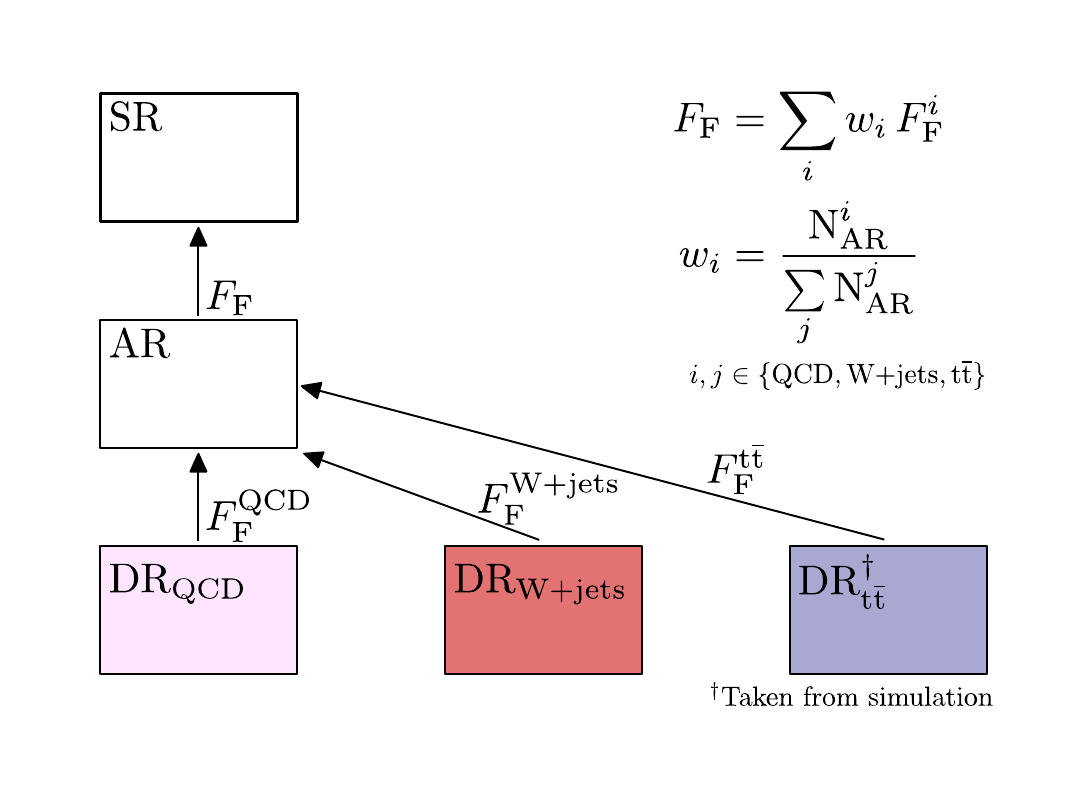}
  \caption {
    Schematic view of the determination and application of the $\FF^{i}$ and $\FF$ for the estimation
    of the background from QCD multijet, $\Wjets$, and $\ttbar$ events due to the misidentification
    of jets as hadronic $\tau$ lepton decays. Note that DR$_{\ttbar}$ is taken from simulation.
  }
  \label{fig:fake-factor-method}
\end{figure}

The underlying assumption in this method is that the ratio of the number of events from background $i$
in the SR to the number of events from the same background in the AR is equal to $N_{\text{pass}}/
N_{\text{fail}}$ in the DR$_i$. This can be ensured by determining $\FF^{i}$ differentially as a
function of several variables taking the most important kinematic or topological dependencies
into account. Residual biases can be removed by adequate corrections, which can be determined from
independent control regions or from simulation. For the analysis the $\FF^{i}$ are estimated from a
fit to the measured values of $\FF^{i}$, as a function of the $\pt$ of the $\tau_{\text{h}}$ candidate
in categories of the $\Pgth$ decay mode, and the jet multiplicity, in bins of $N_{\text{jet}}=0$ or
$N_{\text{jet}}\geq1$. This is in general done in three dedicated and exclusive DR$_{i}$ for the
backgrounds due to QCD multijet, $\Wjets$, and $\ttbar$ events. From the individually determined
$\FF^{i}$ a weighted factor $\FF$ is then obtained on an event-by-event basis from
\begin{linenomath}
  \begin{equation}
    \FF = \sum\limits_{i}w_{i}\,\FF^{i},\qquad
    w_{i} = \frac{N_\mathrm{AR}^{i}}{\sum\limits_{j} N_\mathrm{AR}^{j}},\qquad
    i,j\in\{\text{QCD}, \Wjets, \ttbar\},
  \end{equation}
\end{linenomath}
where $N_\mathrm{AR}^{i}$ corresponds to the expected number of events for background $i$ in the AR.
The factor $\FF$ is then applied to all events in the AR to obtain an estimate for the number and shape
of the sum of QCD multijet, $\Wjets$, and $\ttbar$ events due to $\text{jet}\to\Pgth$ misidentification.
For this purpose the subdominant contributions from $\PZ\to\ell\ell$, diboson and single $\cPqt$ quark events
are subsumed into the $\Wjets$ estimate. The estimates for $N_\mathrm{AR}^{\Wjets}$ and $N_\mathrm{AR
}^{\ttbar}$ are taken from the simulation. The estimate for $N_\mathrm{AR}^{\text{QCD}}$ is obtained
from the events in the AR after subtracting all other backgrounds. These estimates are cross-checked
using a template fit to the data in the AR equivalent to the fit described in
Section~\ref{sec:signal-extraction}, but with the $\pt^{\Pgth}$ distribution as the input shape. From
the resulting distributions, the expected contribution from events with genuine hadronic $\tau$ lepton
decays or electrons or muons misidentified as hadronic $\tau$ lepton decays are subtracted using the
simulation. The principle of the method is outlined in Fig.~\ref{fig:fake-factor-method}. The final
state specific parts of the application of this method in the $\etau$, $\mutau$, and $\tautau$
final states are described in the following.

\subsubsection{Background estimation in the $\etau$ and $\mutau$ final states}

For the $\etau$ and $\mutau$ final states DR$_{\text{QCD}}$ is defined by the same selection as for
the SR, but the electric charges of the $\tau\tau$ pair are required to be of the same sign. To reduce
the contamination from $\Wjets$ events the transverse mass is required to be $\mTem<40\GeV$, and the
relative isolation requirement on the electron (muon) is changed to be $0.05<\Irelem<0.15$ in both final
states. The definition of DR$_{\Wjets}$ also uses the same selection as for the SR, but the requirement
on the transverse mass is changed to $\mTem>70\GeV$ to enrich this background and an additional
requirement of the absence of b jets in the event is imposed to reduce the contamination from $\ttbar$
events. In the $\etau$ and $\mutau$ final states $\ttbar$ production is a subdominant background
with respect to $\Wjets$ and QCD multijet events. Since there is no sufficiently populated pure DR
for $\ttbar$ events covering a similar phase space as the SR, the $\FF^{\ttbar}$ are estimated from
simulation after the event selection and before the event categorization.

\subsubsection{Background estimation in the $\tautau$ final state}

The $\tautau$ final state deviates in two aspects from the $\etau$ and $\mutau$ final states. Firstly,
QCD multijet production is by far the dominant background. Therefore only DR$_\text{QCD}$ is defined
from the single requirement that the electric charges of the $\tau\tau$ pair should be of the same
sign. The $\FF^{\text{QCD}}$ are then also used to estimate the background from $\Wjets$ and $\ttbar$
events. Secondly, misidentified $\Pgth$ candidates from QCD multijet production usually originate
from $\text{jet}\to\Pgth$ misidentification. In this way a combinatorial effect arises for the
determination of $\FF^{\text{QCD}}$ from the fact that each event can enter the AR in one of two
mutually exclusive ways, either if the leading $\Pgth$ candidate fulfills the nominal $\Pgth$
identification requirement and the subleading $\Pgth$ candidate the inverted requirement or vice versa.
This combinatorial effect is taken into account by assigning a weight of 0.5 to these types of events.
For the backgrounds from $\Wjets$ and $\ttbar$ events typically only one of the reconstructed $\Pgth$
candidates originates from a misidentified jet and the other one from a genuine $\tau$ lepton decay.
The fraction of events with two misidentified jets is at most a few percent and thus well below the
associated systematic uncertainties. Since there are no significant combinatorial effects involved,
these events are considered with a weight of~$1$.

\subsubsection{Corrections to the $\FF^{i}$ in the $\etau$, $\mutau$, and $\tautau$ final states}

For each of the backgrounds considered, corrections to the estimated $\FF^{i}$ are determined in modified
determination regions to account for residual biases of the method. An overview of these corrections
is given in Table~\ref{tab:ff-corrections}.

As described above the $\FF^{i}$ are obtained from fits of a functional form to the estimated values
of the ratio given in Eq.~(\ref{eq:fake-factor}) in bins of the $\Pgth$ candidate $\pt$. Additional
dependencies on the $\Pgth$ decay mode and the jet multiplicity are taken into account.
The choices of the functional forms that are fitted to the data, the finite binning in the
$\Pgth$ candidate $\pt$, and the omission of further, potentially important, dependencies on kinematic
or topological variables may lead to such biases. These effects are checked and corrected for, for
each of the $\FF^{i}$ individually, in the DR$_{i}$ themselves by comparing the actual number of events
with the $\Pgth$ candidate matching the \text{Tight/Medium} working point of the $\Pgth$ identification
discriminant to the number of events estimated from the method. Residual corrections are determined
as a function of the invariant mass of the visible decay products of the $\tau\tau$ pair, $\mvis$,
and found to be compatible with unity within the statistical precision. This demonstrates that the
main dependencies of the $\FF^{i}$ are taken into account. In Table~\ref{tab:ff-corrections} these
corrections are labeled as ``Nonclosure''.

For $\FF^{\text{QCD}}$ two additional corrections are applied: in the $\etau$ ($\mutau$) final state
a correction (``$\Irelem$-dependent'') is obtained as a function of $\Irelem$ by comparing the number
of events matching the \text{Tight/Medium} working point of the $\Pgth$ identification discriminant
to the number of events estimated from the method in a control region equivalent to DR$_{\text{QCD}}$,
with the only difference being that the initial requirement on $\Irelem$ is dropped. This correction
is found to be $\mathcal{O}(10\%)$ and compatible with unity within one standard deviation of the
statistical precision. In the $\tautau$ final state a correction (``$\pt^{\Pgth}$-dependent'') is
derived as a function of the $\pt$ of the other $\Pgth$ candidate. This correction is found to range
between a few percent and $20\%$.

\begin{table}[b]
  \topcaption{
    Corrections applied to the $\FF^{\text{QCD}}$, $\FF^{\Wjets}$, and $\FF^{\ttbar}$ as described in
    the text. In the fourth column the source is indicated from which the correction is derived. The
    dependency $\pt^{\Pgth}$ in the third line refers to the $\pt$ of the $\Pgth$ candidate that is
    assumed to originate from a genuine $\tau$ lepton decay.
  }
  \label{tab:ff-corrections}
  \centering
  \begin{tabular}{llclccc}
    & Correction & Dependency & Source & $e\Pgth$ & $\mu\Pgth$ & $\Pgth\Pgth$ \\
    \hline
    & Nonclosure & $\mvis$ & DR$_{\text{QCD}}$ & $\checkmark$ & $\checkmark$ & $\checkmark$ \\[\cmsTabSkip]
    \multirow{2}{*}{$\FF^{\text{QCD}}\vphantom{^{\Pe(\mu)}}$} & \multirow{2}{*}{$\Irelem$-,
    $\pt^{\Pgth}$-dependent} & $\Irelem$ & DR$_{\text{QCD}}$
    (w/o $\Irelem$) & $\checkmark$ & $\checkmark$ & \\
    & & $\pt^{\Pgth\dagger}$ & DR$_{\text{QCD}}$  & & & $\checkmark$ \\
    & Opposite/Same charge & $\mvis$ & Orthogonal iso./ID$^{\dagger}$ & $\checkmark$ & $\checkmark$ & $\checkmark$ \\[\cmsTabSkip]
    \multirow{2}{*}{$\FF^{\text{W+jets}}$} & Nonclosure & $\mvis$ & DR$_{\Wjets}$ & $\checkmark$ & $\checkmark$ &  \\
    & $\mTem$-dependent & $\mTem$ & From simulation & $\checkmark$ & $\checkmark$ &  \\[\cmsTabSkip]
    \multirow{2}{*}{$\FF^{\text{t}\bar{\text{t}}}$} & Nonclosure & $\mvis$ & DR$_{\ttbar}$ & $\checkmark$ & $\checkmark$ &  \\
    & Data/Simulation & None & $\text{t}\bar{\text{t}}$ enriched sideband & $\checkmark$ & $\checkmark$ &  \\[\cmsTabSkip]
    \hline
    \multicolumn{7}{l}{$^{\dagger}$ \small{Refers to the $\Pgth$ candidate that is assumed to originate from a genuine
    $\tau$ lepton decay.}} \\
  \end{tabular}
\end{table}

For all final states another correction (``Opposite/Same charge'') is derived to account for the
transition from DR$_{\text{QCD}}$ with the same charge requirement on the $\tau\tau$ pair to the SR
with an opposite charge requirement. This correction is determined as a function of $\mvis$ in a
control region with $0.1<I_\text{rel}^{\Pe}<0.2$ ($0.15<I_\text{rel}^{\mu}<0.25$) in the $\etau$
($\mutau$) final state and in a control region where the other $\Pgth$ candidate matches the
\text{VeryLoose} but fails the \text{Medium} working point of the $\Pgth$ identification discriminant
in the $\tautau$ final state. In all final states the correction is found to be compatible with unity
within one standard deviation of the statistical precision, which ranges from 10 to 20\% in the $\etau$
and $\mutau$ final states, respectively, and from a few percent to 10\% in the $\tautau$ final state.

In the $\etau$ ($\mutau$) final state two more corrections are applied. Firstly, for the $\FF^{\Wjets}$
a residual dependence is expected from the selection requirements on $\ptem$: for low $\mTem$ a value
of $\ptem$ above the thresholds of the offline selection will lead to a harder hadronic recoil and
more jets in the event. This in turn may lead to less isolated $\Pgth$ candidates especially at low
$\pt^{\Pgth}$. A correction (``$\mTem$-dependent'') for this effect as a function of $\mTem$ is
derived from simulation. It ranges from 10 to 30\%, while usually compatible with unity within one
standard deviation of the statistical precision. It is assumed to be the same for $\PZ\to\ell\ell$,
diboson, and single $\cPqt$ quark events. Secondly, as described above, the $\FF^{\ttbar}$ are obtained from
simulation. Data-to-simulation corrections are derived from a control region in data, which is
characterized by the presence of at least one b jet and at least one lepton pair consisting of an
isolated electron and an isolated muon in the event. Since this correction is found to be independent
of $\ptem$, $\mTem$, or $\mvis$, within the experimental precision, a common factor is used depending
on the final state ($\etau$ or $\mutau$) and the $\Pgth$ decay mode.

\subsubsection{Background estimation in the $\emu$ final state}

In the $\emu$ final state the background from QCD multijet events is estimated from an AR fulfilling
the same selection requirements as the SR, however the charges of the leptons are required to be of
the same sign. Extrapolation factors for the same charge to the opposite charge phase space are
obtained in bins of the $\pt$ of the two leptons and their separation in $\Delta R$. These extrapolation
factors are derived in a DR without event categorization, in which the isolation requirements on the
leptons are chosen to be orthogonal to the SR. Finally, corrections are applied to account for the
extrapolation into the exclusive event categories and for the extrapolation into the SR. The corrections
for the extrapolation into the exclusive event categories are determined from the same DR, but inclusive
in the $\pt$ of, and separation between, the leptons. They are about 0.6 (1) for all \text{b-tag}
(\text{no b-tag}) categories. The correction for the extrapolation into the SR is about 0.9 as
determined from simulation.

\subsection{Backgrounds estimated from simulation}
\label{subsec:MC-backgrounds}

All other backgrounds, apart from the ones described in Section~\ref{subsec:Data-backgrounds} are
estimated from simulation. Corrections are derived to account for residual differences in the efficiency
of the selected trigger paths, in the electron and muon tracking efficiency, and in the efficiency of
the identification and isolation requirements, for electrons and muons. These corrections are obtained
using the ``tag-and-probe'' method, as described in Ref.~\cite{Khachatryan:2010xn}, with $\PZ\to\ee$
and $\PZ\to\mumu$ events in bins of \pt and $\eta$ of the probed electron or muon. They are usually
not larger than a few percent. In a similar way, corrections are obtained for the efficiency of
triggering on the hadronic $\tau$ lepton decays in the $\tautau$ final state and for the $\Pgth$
identification efficiency. In this case the tag-and-probe method is applied to $\PZ\to\tau\tau$ events
in the $\mutau$ final state.

The energy of a jet is corrected to the expected response of the jet at stable-hadron level, using
corrections measured in bins of the jet \pt and $\eta$. These corrections are usually not larger than
10 to 15\%. Residual data-to-simulation corrections are applied to the simulated samples. They
usually range between sub-percent level at high jet \pt in the central part of the detector to a
few percent in the forward region. A correction is applied to the direction and magnitude of the
$\ptvecmiss$ vector based on the differences between the estimates of the hadronic recoil in $\PZ\to
\mumu$ events in data and simulation. This correction is applied to $\PZ\to\tau\tau$, $\Wjets$, and
signal events, where a well-defined direction and magnitude of genuine $\ptvecmiss$ can be defined. The
efficiency for genuine and misidentified b jets to pass the \text{Medium} working point of the b
tagging discriminator is determined from data, using $\ttbar$ events for genuine b jets and $\PZ
\text{+jets}$ events for jets predominantly originating from light-flavor quarks. Data-to-simulation
corrections are obtained for these efficiencies and used to correct the number of b jets in the
simulation, which translates into the number of events in the global \text{b-tag} and \text{no b-tag}
event categories. In the $\emu$ final state data-to-simulation corrections are derived for the rate
at which jets are misidentified as an electron or muon. These are determined as a function of the jet
$\pt$ from $\PZ\text{+jets}$ events in the $\PZ\to\ell\ell$ decay. They are applied to $\Wjets$ and
diboson events, which form more than 90\% of the expected background due to $\text{jet}\to\ell$
misidentification in the $\emu$ final state, and where the flavor composition of jets is similar to
that in the region in which the corrections are determined. Corrections are further applied to $\PZ
\to\mumu$ events in the $\mutau$ and $\tautau$ final states in which a muon is reconstructed as a
$\Pgth$ candidate and in $\PZ\to\ee$ events in the $\etau$ and $\tautau$ final states in which an
electron is reconstructed as a $\Pgth$ candidate, to account for residual differences in the $\ell
\to\Pgth$ misidentification rate between data and simulation. Finally a correction, obtained from
$\PZ\to\ee$ events, to the energy scale for electrons misidentified as hadronic $\tau$ lepton decays
is applied. Corresponding uncertainties in all these corrections are incorporated into the uncertainty
model discussed in Section~\ref{sec:uncertainties}.

Deficiencies in the modeling of Drell--Yan events in the $\Pe\Pe$, $\mu\mu$ and $\tau\tau$
final states are corrected for by a weighting of the simulated $\PZ\to\mumu$ events to data in bins
of $\pt(\mu\mu)$ and $m(\mu\mu)$. The weights obtained are applied to the simulated events in all
leptonic final states. For the statistical analysis the overall normalization of the background from
$\PZ\to\tau\tau$ events is furthermore constrained by dedicated control regions of $\PZ\to\mumu$
events in each global event category, making use of the equal branching fractions for the $\PZ$ boson
decays into $\tau$ leptons or muons, in the context of lepton universality. Theoretical uncertainties
arising from residual kinematic differences between the selected dimuon and $\tau\tau$ final states
are incorporated into the uncertainty model. In addition all simulated $\ttbar$ events are weighted
to better match the top quark \pt distribution, as observed in data~\cite{Khachatryan:2015oqa}.
For the statistical analysis the overall normalization of this background is also constrained by a
dedicated control region with an isolated electron, an isolated muon, and large $\ptmiss$ in the
final state, which is chosen to be orthogonal to the SR in the $\emu$ final state; this sample has
a $\ttbar$ purity of $85\%$. All control regions used for the statistical analysis are outlined in
Fig.~\ref{fig:categories}.

\subsection{Cross-checks of background estimations}
\label{subsec:cross-checks}

Two cross-checks are performed to give confidence in the background estimation. In a first cross-check
all backgrounds apart from QCD multijet production and the normalization for $\Wjets$ events are taken
from simulation. For this purpose all corrections as summarized in Section~\ref{subsec:MC-backgrounds}
are applied to all simulated events. This cross-check is performed in the $\etau$ and $\mutau$ final
states individually.

The $\Wjets$ prediction, prior to the statistical inference of the signal, is obtained by subtracting
the small contribution of all other backgrounds except for QCD multijet and $\Wjets$ events from data
in corresponding control regions requiring the charges of the $\tau\tau$ pair to be of opposite (OS)
or same sign (SS) and $\mTem>70\GeV$.  An estimate for the normalization of the QCD multijet and $\Wjets$
events can then be obtained from the following system of linear equations
\begin{linenomath}
  \begin{equation}
    \begin{split}
      N_{\text{data}}^{\prime \text{SS}} &= \hphantom{f_{\text{QCD}}^{\text{OS/SS}}}\,
      N_{\text{QCD}}^{\text{SS}} + \hphantom{f_{\Wjets}^{\text{OS/SS}}}\,
      N_{\Wjets}^{\text{SS}} \\
      N_{\text{data}}^{\prime \text{OS}} &= f_{\text{QCD}}^{\text{OS/SS}}\, N_{\text{QCD}}^{\text{SS}}
      + f_{\Wjets}^{\text{OS/SS}}
      \, N_{\Wjets}^{\text{SS}},\\
    \end{split}
    \label{eq:NQCD-estimate}
  \end{equation}
\end{linenomath}
where $N_{\text{data}}^{\prime \text{SS(OS)}}$ corresponds to the number of events in the control
regions, after subtracting the expected number of events for all other backgrounds, and $f_{\text{QCD}
(\Wjets)}^{\text{OS/SS}}$ is the expected OS to SS ratio for $\Wjets$ and QCD multijet events. For
this estimate $f_{\Wjets}^{\text{OS/SS}}$ is obtained from the simulation and $f_{\text{QCD}}^{
\text{OS/SS}}$ from another control region with inverted isolation requirements on the electron or
muon, as described below. An estimate for $N_{\Wjets}^{\text{SS}}$ can then be obtained from
Eq.~(\ref{eq:NQCD-estimate}). From this the number of $\Wjets$ events in the SR can be inferred via
$f_{\Wjets}^{\text{OS/SS}}$ and another extrapolation factor from the control region into the SR,
which again is taken from simulation. To stay as close as possible to the kinematic regime in the
signal regions an OS and an SS control region for the determination of $N_{\text{data}}^{\prime
\text{OS}}$ and $N_{\text{data}}^{\prime \text{SS}}$ is defined, for each event subcategory in the
$\etau$ and $\mutau$ final states, as described in Section~\ref{sec:event-selection}, amounting
to eight control regions per final state. The shape of the final discriminating variable distribution
used for the signal extraction is taken from simulation.

The shape and normalization of the QCD multijet background distributions prior to the signal extraction
are obtained from control regions equivalent to the signal regions with the exception of a SS instead
of an OS requirement on the charge of the selected $\tau\tau$ pair. From the events in this control
region all other expected backgrounds are subtracted using the normalization and shape information for
the final discriminating variable distribution from simulation, with the exception of the normalization
of $\Wjets$ events, which is obtained as described above. The extrapolation factors ($f_{\text{QCD}}^{
\text{OS/SS}}$) from the SS to OS selection are obtained from control regions, where in addition, to
the corresponding charge requirement, the isolation requirement on the electron or muon is inverted.
The extrapolation factors are then obtained from a fit to the data in the control regions similar to
the one described in Section~\ref{sec:signal-extraction}. To control the normalization of the $\Wjets$
and QCD multijet events the eight additional control regions per final state, as introduced above, are
added to the fit for the signal extraction and the corresponding normalization uncertainties are
incorporated into a modified uncertainty model.

In a second cross-check the background from $\PZ\to\tau\tau$ events in the main analysis is replaced
with the prediction obtained from the $\mu\to\tau$ \emph{embedding} method as used during the LHC Run-1
analyses and described, for example, in Refs.~\cite{Chatrchyan:2014nva,Aad:2015kxa}. In this process
$\PZ\to\mumu$ events are selected in data. The muons are then replaced by simulated $\tau$ lepton
decays with the same kinematic properties as the reconstructed muons. In this way the method relies 
only on the simulation of the well understood $\tau$ lepton decay while all other parts of the event 
are obtained from data. As a consequence several data-to-simulation corrections as described in
Section~\ref{subsec:MC-backgrounds}, which are of particular importance for the event categorization
as well as for the shape of the final discriminating variable distribution, do not need to be applied
for this process. This applies, for example, to corrections of the jet energy scale, b tagging efficiency,
and $\ptvecmiss$. This cross-check is applied in the $\etau$, $\mutau$, and $\tautau$ final states
individually. Both the extrapolation factors from the inclusive event selection into the event
subcategories, as well as the shapes of the final discriminating variable distribution for the signal
extraction, as obtained from the simulation, are found to be in good agreement with the estimates as
obtained from the embedding method, within the estimated uncertainties. In addition the uncertainties
that are related to the experimental aspects of the $\mu\to\tau$ embedding, which are orthogonal to
the uncertainties in the estimate from simulation, are incorporated into a modified uncertainty model
to replace several uncertainties for the estimate based on the simulation.

\section{Statistical inference for the signal}
\label{sec:signal-extraction}

The final discriminating variable used to search for a signal is the total transverse mass,
$\mTtot$~\cite{Aad:2014vgg}, defined as
\begin{linenomath}
  \begin{equation}
    \mTtot = \sqrt{\mT^{2}(\pt^{\tau_{1}},\pt^{\tau_{2}})+\mT^{2}(\pt^{\tau_{1}},\ptmiss)+\mT^{2}
      (\pt^{\tau_{2}},\ptmiss)},
    \label{eq:mttot}
  \end{equation}
\end{linenomath}
where the pair $(\tau_{1},\tau_{2})$ can be $(\Pe,\mu)$, $(\Pe,\Pgth)$, $(\mu,\Pgth)$, or
$(\Pgth,\Pgth^{\prime})$, and the transverse mass, $\mT$, between two objects with transverse momenta
$\pt$ and $\pt^{\prime}$, and relative difference $\Delta\phi$ in the azimuthal angle is given by:
\begin{linenomath}
  \begin{equation}
    \mT = \sqrt{2\,\pt\,\pt^{\prime}\left[1-\cos(\Delta\phi)\right]}.
  \end{equation}
\end{linenomath}

\begin{figure}[htbp]
  \centering
  \includegraphics[width=.48\textwidth]{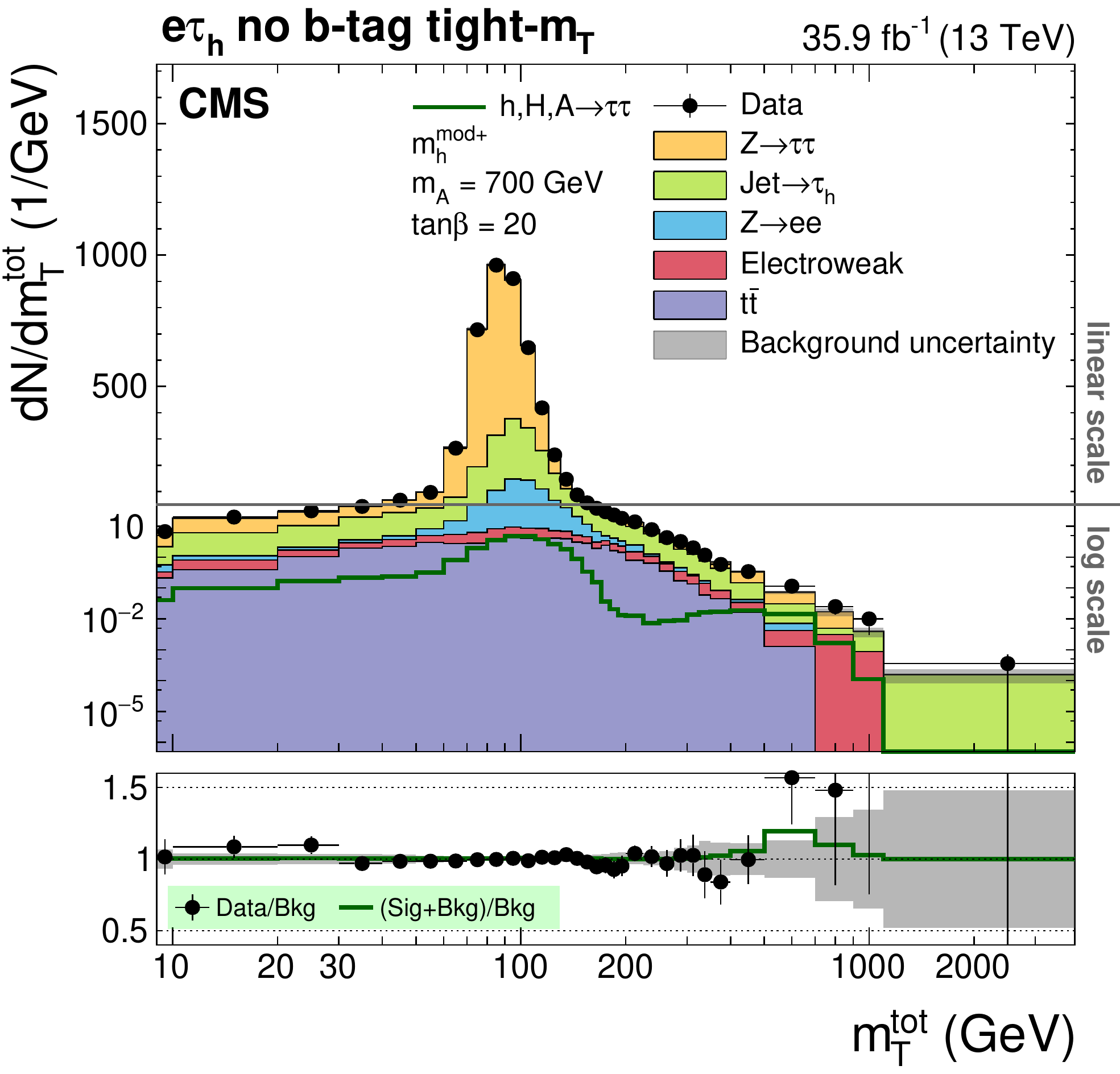}
  \includegraphics[width=.48\textwidth]{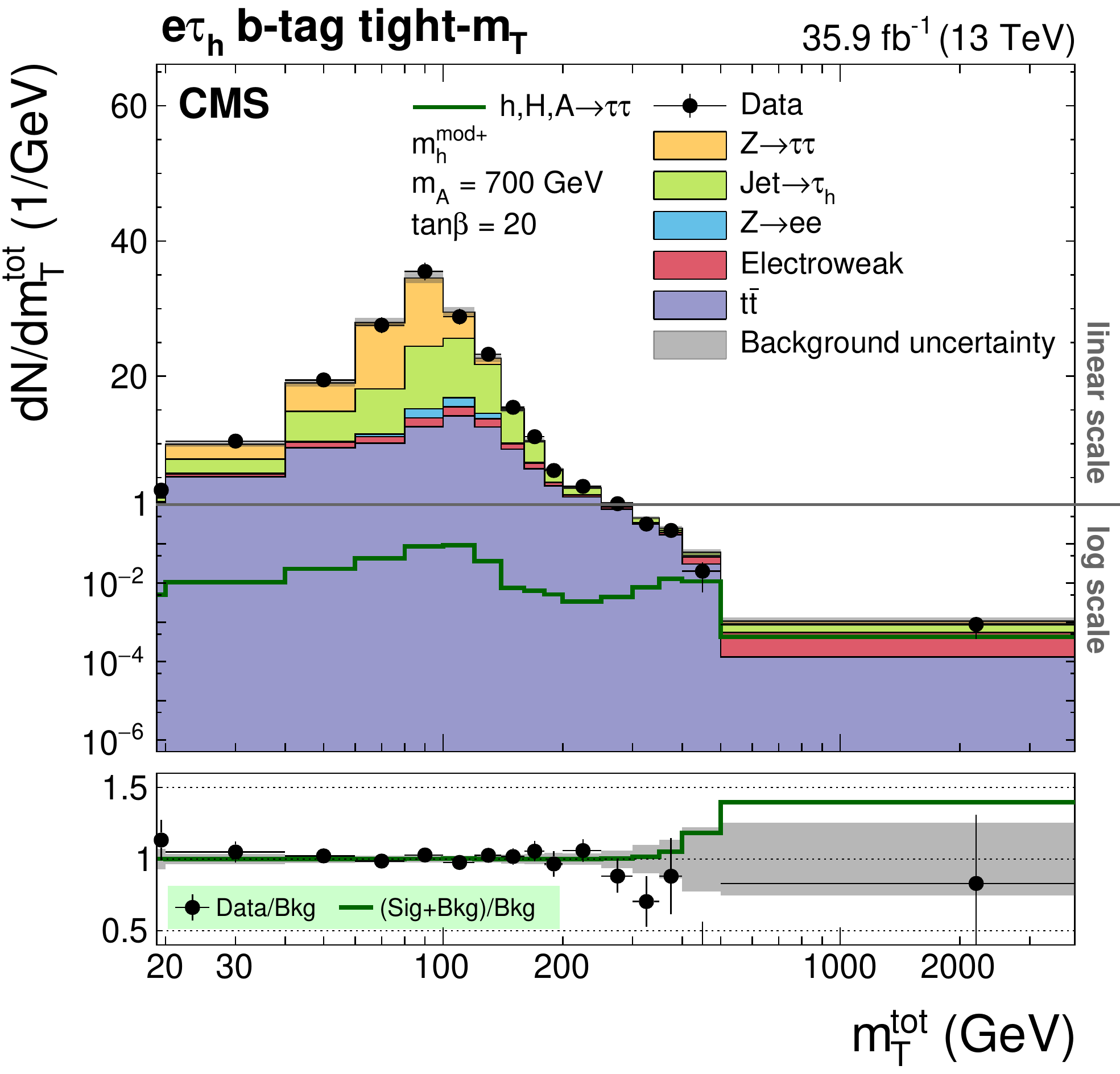}
  \includegraphics[width=.48\textwidth]{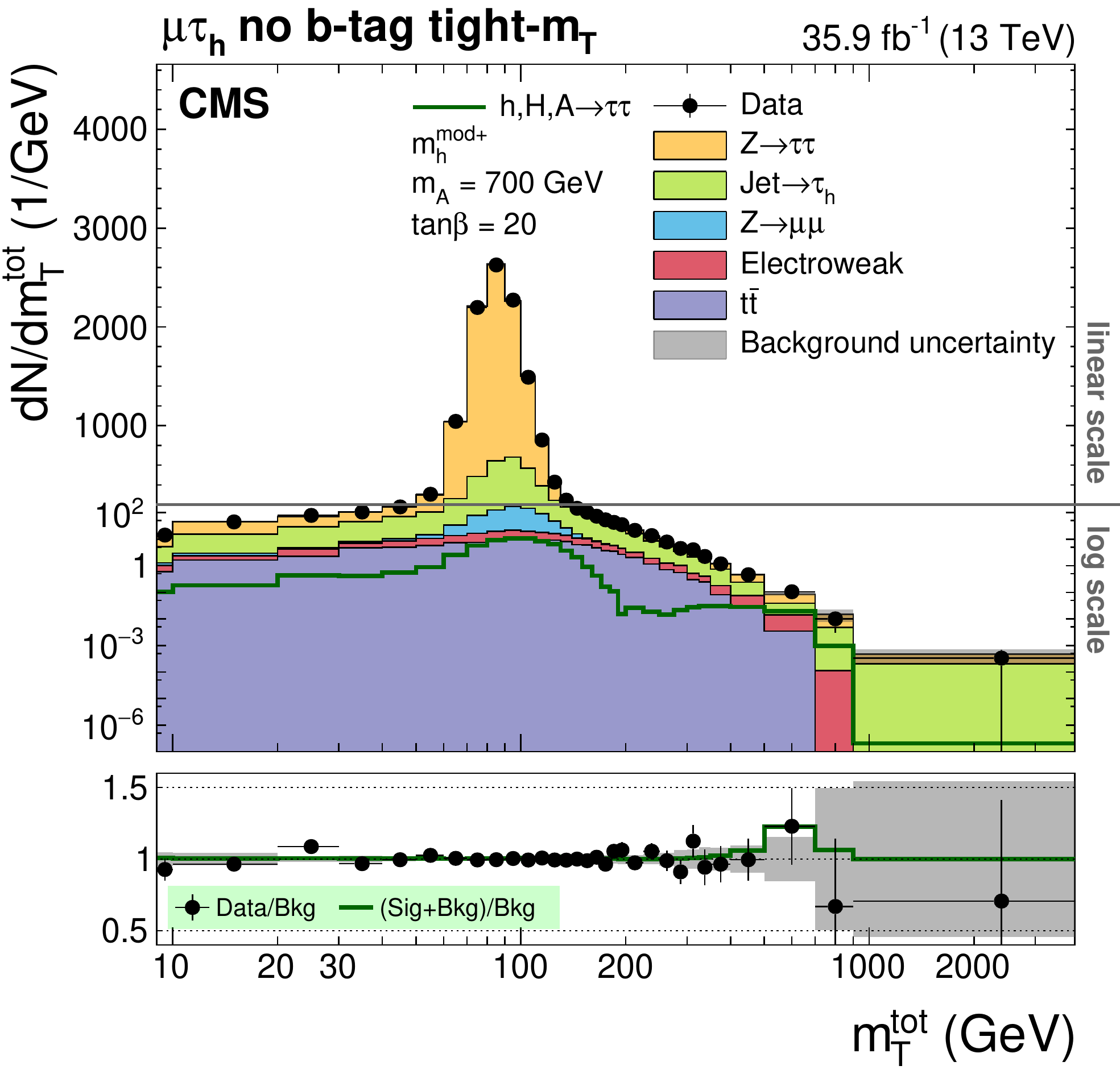}
  \includegraphics[width=.48\textwidth]{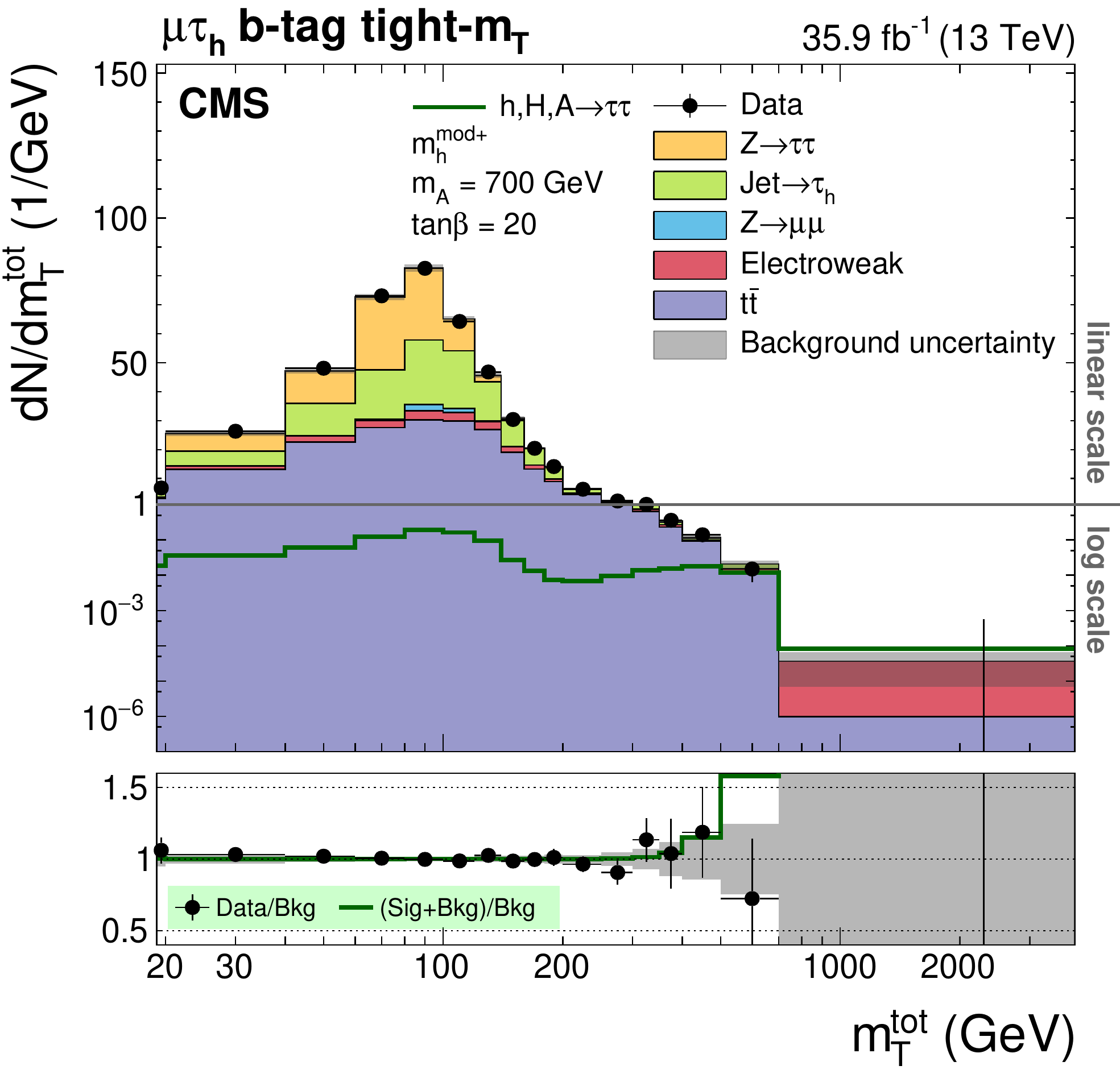}
  \caption {
    Distribution of $\mTtot$ in the global \text{no b-tag} (left) and \text{b-tag} (right) categories
    in the $\etau$ (upper row) and $\mutau$ (lower row) final states. In all cases the most sensitive
    \text{tight-}$\mT$ event subcategory is shown. The gray horizontal line in the upper panel of each
    subfigure indicates the change from logarithmic to linear scale on the vertical axis.
  }
  \label{fig:mTtot-distributions-1}
\end{figure}

\begin{figure}[htbp]
  \centering
  \includegraphics[width=.48\textwidth]{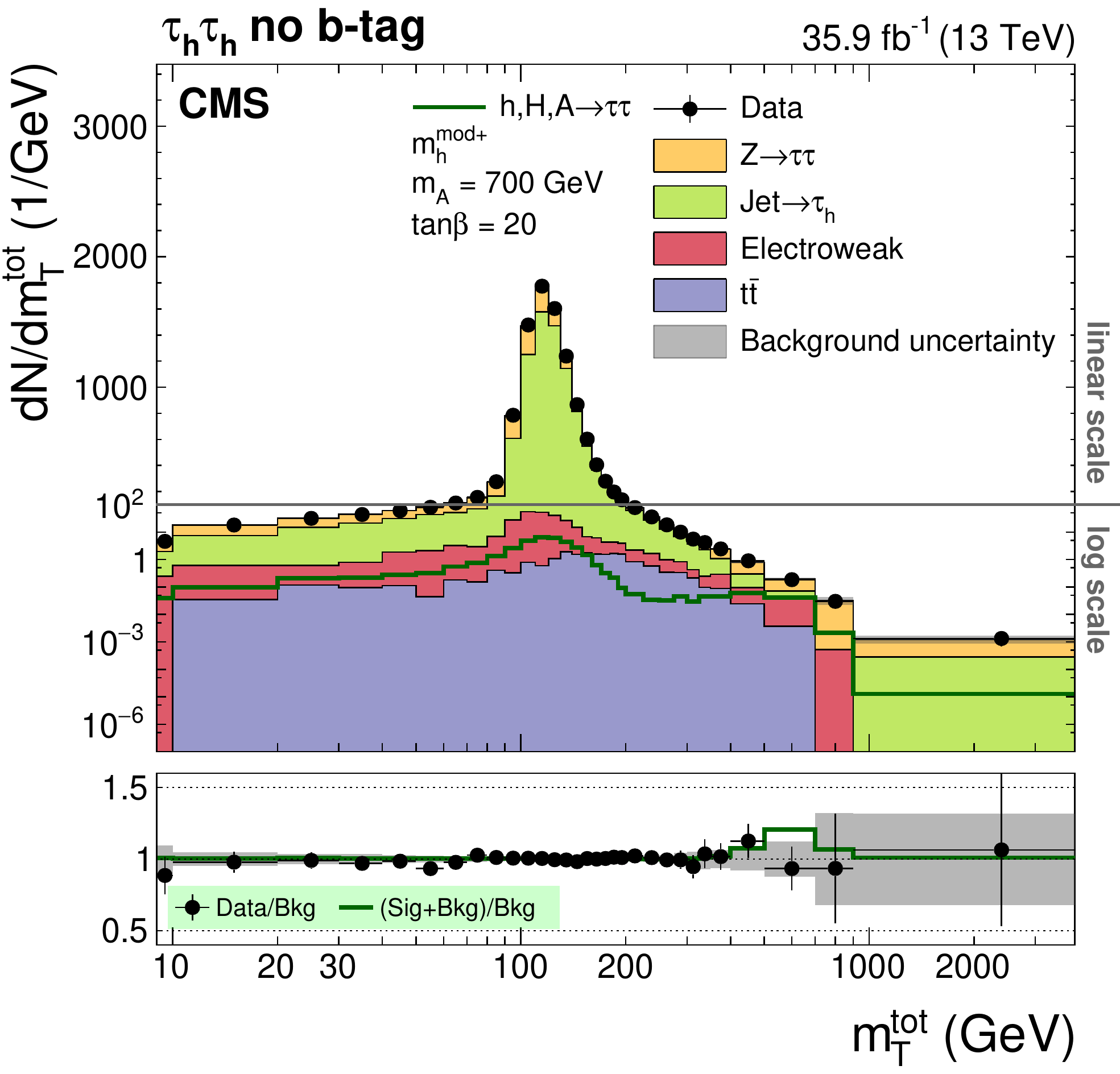}
  \includegraphics[width=.48\textwidth]{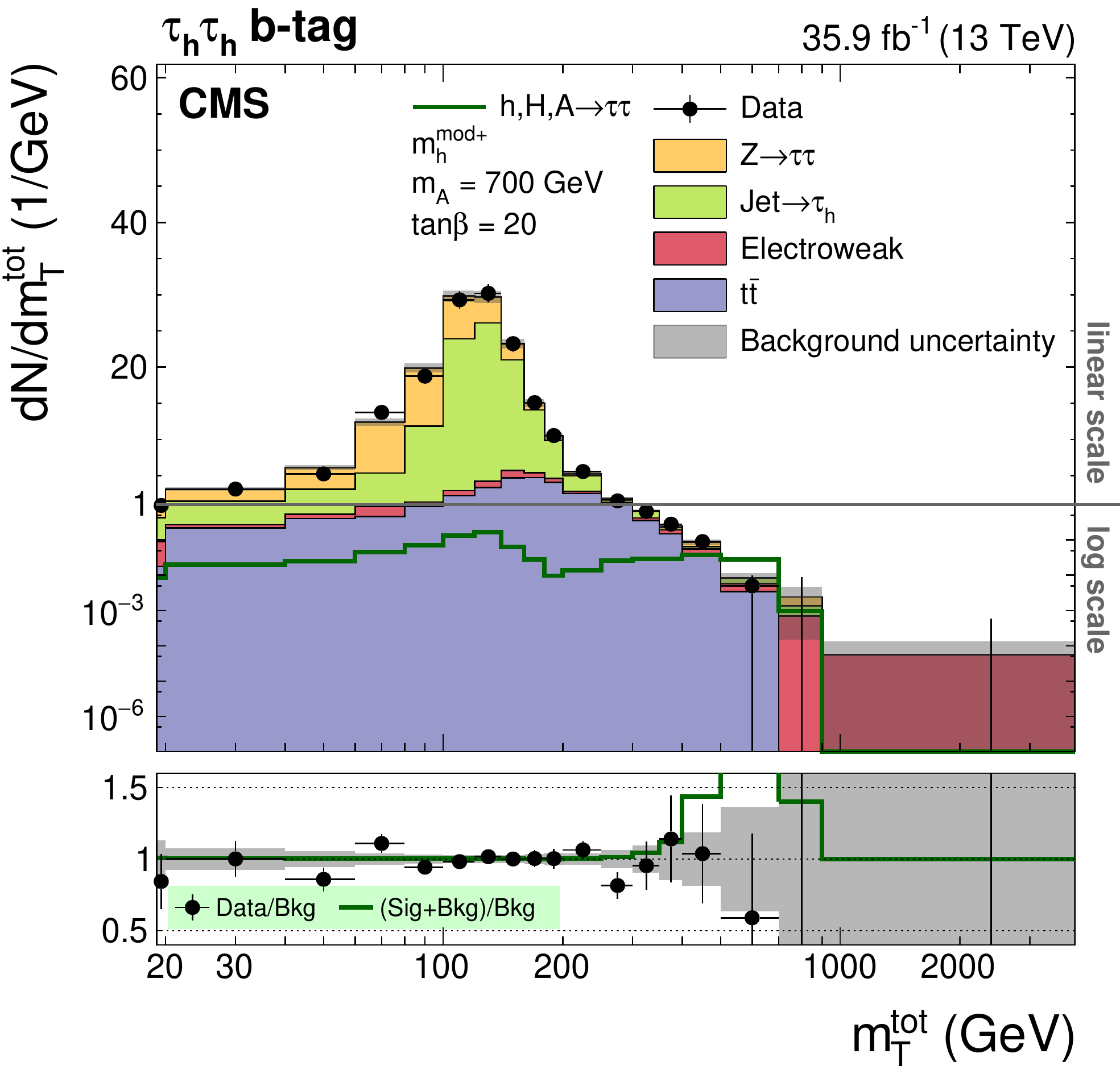}
  \includegraphics[width=.48\textwidth]{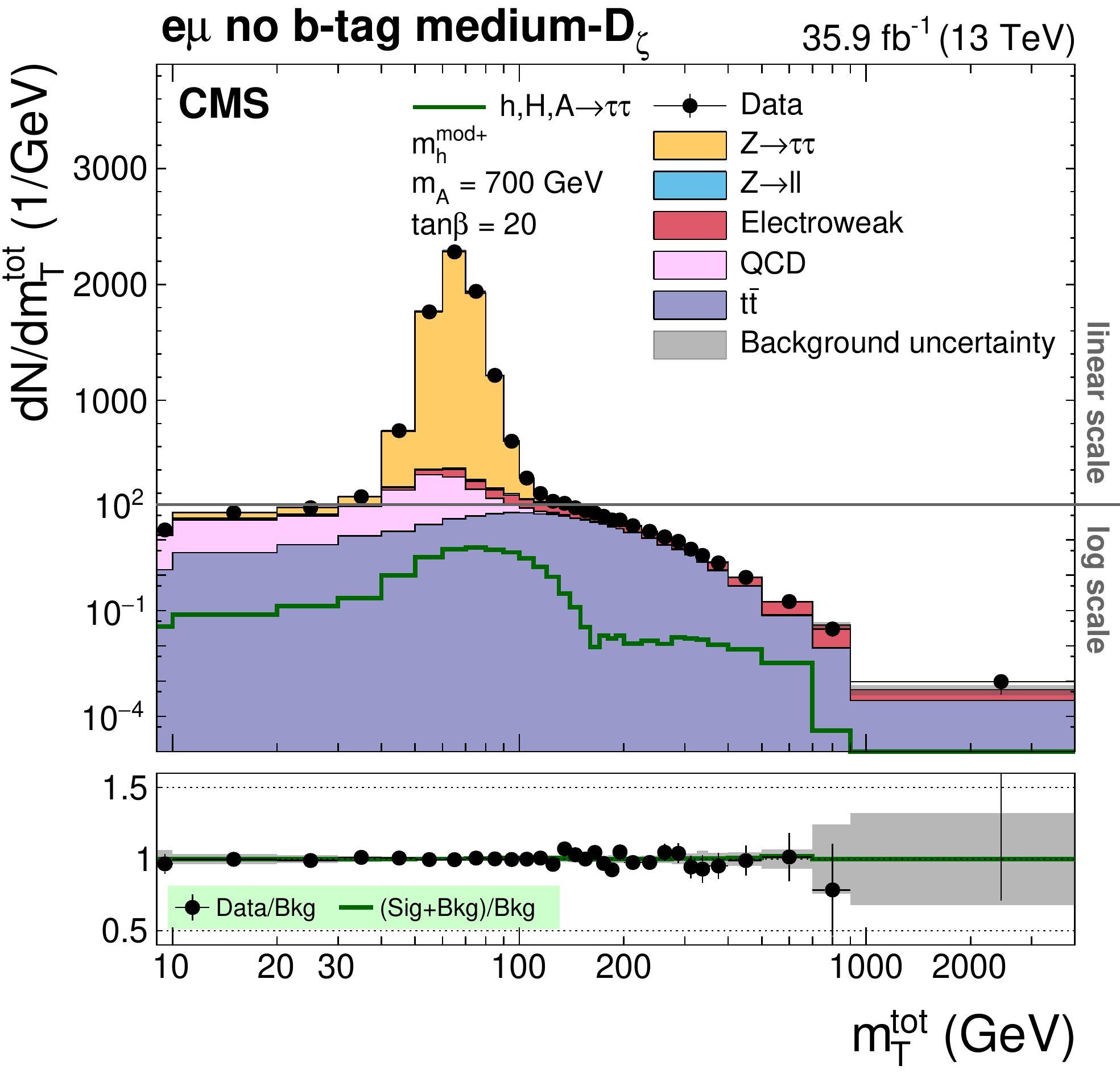}
  \includegraphics[width=.48\textwidth]{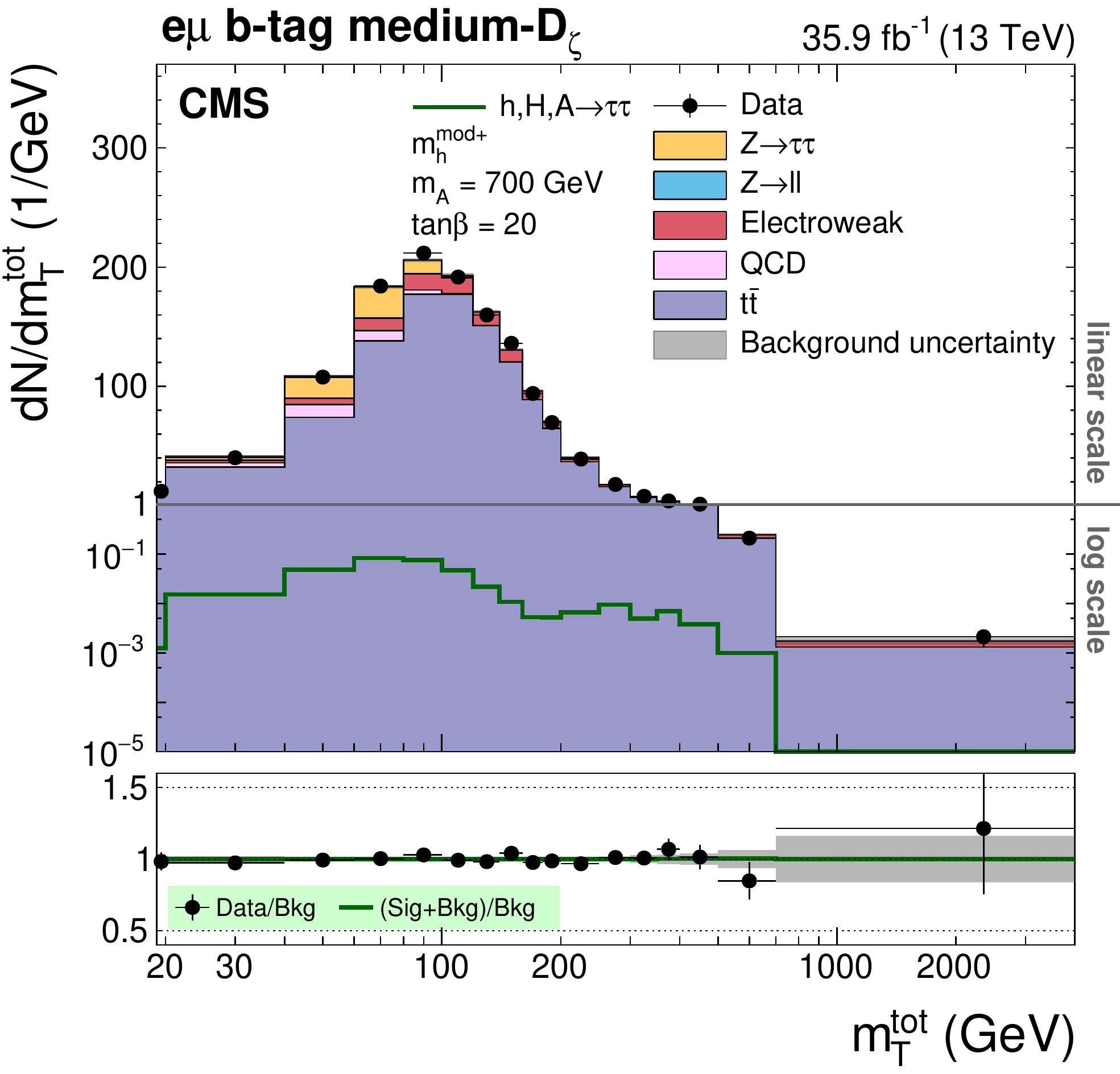}
  \caption {
    Distribution of $\mTtot$ in the global \text{no b-tag} (left) and \text{b-tag} (right) categories
    in the $\tautau$ (upper row) and $\emu$ (lower row) final states. For the $\emu$ final state the
    most sensitive \text{medium-}$\Dzeta$ event subcategory is shown. The gray horizontal line in the
    upper panel of each subfigure indicates the change from logarithmic to linear scale on the vertical
    axis.
  }
  \label{fig:mTtot-distributions-2}
\end{figure}

The input distributions to the statistical inference of the signal in a subset of the most sensitive
event subcategories per final state are shown in Figs.~\ref{fig:mTtot-distributions-1}
and~\ref{fig:mTtot-distributions-2}. The expected $\mTtot$ distribution is represented by the stack
of filled histograms in the upper panel of each subfigure, where each filled histogram corresponds
to the estimated template distribution of the given SM process that has been taken into account for
the analysis. For this purpose the fractions of QCD multijet, $\Wjets$, and $\ttbar$ events
contributing to the event selection by $\text{jet}\to\Pgth$ misidentification are subsumed
into one single contribution labeled as ``$\text{jet}\to\Pgth$''. The remaining fractions
from $\Wjets$, single $\cPqt$ quark, and diboson events are subsumed into one single contribution labeled
as ``Electroweak''. The shaded band associated with the sum of filled histograms corresponds to the
combination of all normalization and shape altering uncertainties in all background processes, taking
into account all correlations as obtained from the fit used for the signal extraction. The ratio of
the data points to the expectation from the sum of all filled histograms is shown in the lower panel
of each subfigure; the statistical uncertainty in the data is represented by the error bars and the
uncertainty in all background processes by the shaded band. The expected $\mTtot$ distribution for a
signal of three neutral Higgs bosons from gluon fusion and the production in association with b quarks
in the MSSM $\mhmodp$ scenario, discussed in Ref.~\cite{Carena:2013ytb}, for $\mA=700\GeV$ and $\tan
\beta=20$ is also shown. The signal distribution reveals two peaking structures, related to the signal
from the h at about 130\GeV, and the nearly mass degenerate H and A at 700\GeV.

To quantify the amount of signal a simultaneous binned maximum likelihood fit to the $\mTtot$
distributions in all event subcategories and all final states is performed. This is done under the
background-only and several signal-plus-background hypotheses to search for potential excesses due
to the presence of additional Higgs bosons over the known SM processes. For this purpose the SM
Higgs boson is included in the background processes. The control regions, which have been designed
to constrain the background from Drell--Yan and $\ttbar$ events, are included in the likelihood
model, resulting in a fit in sixteen event subcategories and three control regions, as outlined in
Fig.~\ref{fig:categories}. To check the validity of the statistical model, prior to this fit, several
goodness of fit tests, based on the background-only hypothesis, have been performed on the input
distributions in each event subcategory. Tests have been chosen, which are sensitive to both kinds
of deviations from the applied model, local deviations in individual bins of the input distribution,
and deviations across several correlated bins, like systematic shifts. All uncertainties and their
correlations have been taken into account for these tests. All tests have indicated good statistical
compatibility. The modeling of important input variables has been checked in control regions, and the
sensitivity and influence of each individual event (sub-)category on the combined result have been
verified, using pseudo-experiments.

The data are interpreted in two ways based on the ratio of the fitted likelihoods for the back\-ground-only
and the tested signal-plus-background hypotheses. For each interpretation the model for the background
processes is formed from the template distributions as shown, for example, in Figs.~\ref{fig:mTtot-distributions-1}
and~\ref{fig:mTtot-distributions-2}. In a first interpretation, which is meant to be as model-independent
as possible, the signal model corresponds to a single resonance, $\phi$, with a width negligible
compared to the experimental resolution. For this purpose, 28 simulated single narrow resonances with
mass $m_{\phi}$ between 90\GeV and 3.2\TeV in the gluon fusion and in association with b quarks are
used. For both production modes the \pt spectrum of the $\phi$ is simulated at NLO precision
as described in Section~\ref{subsec:simulation}. The signal is searched for in both production modes
at the same time, using two freely varying parameters of interest for the fit to the data, one for
each production mode. In a second interpretation, the simulated mass points are combined into
the multiresonance signal structure expected from each of the tested MSSM benchmark scenarios.
This is done using the model predictions, as described in Sections~\ref{subsec:simulation}
and~\ref{sec:results}, and a linear template morphing algorithm, as described in Ref.~\cite{Read:1999kh},
to move the simulated mass points to their exact predicted values.

\section{Systematic uncertainties}
\label{sec:uncertainties}

The uncertainty model comprises theoretical uncertainties, experimental uncertainties, and uncertainties
due to the limited population of the template distributions used for the prediction of the background
processes. The last group of uncertainties are most important for the high-mass Higgs boson searches.
All systematic uncertainties are implemented in the form of nuisance parameters in the likelihood,
which can be further constrained by the fit to the data. The following uncertainties are implemented
as normalization uncertainties that leave the shape of the $\mTtot$ distributions unchanged:

\begin{itemize}
\item
The uncertainty in the integrated luminosity measurement is 2.5\%~\cite{CMS:2017sdi}. It is applied
to all processes that have been estimated from simulation.
\item
The uncertainties in the measurement of the identification, isolation, and trigger efficiencies are
found to amount to 2\% both for electrons and muons, adding all individual contributions in quadrature.
These uncertainties are applied to all processes that are estimated from simulation.
\item
Uncertainties in the measurement of the probability of electrons ($\Pe\to\Pgth$) and muons ($\Pgm\to
\Pgth$) to be misidentified as hadronic $\tau$ lepton decays are applied to the fraction of simulated
Drell--Yan events with light leptons being misidentified as hadronic $\tau$ lepton decays in the
$\etau$, $\mutau$, and $\tautau$ final states. The uncertainty in the $\Pe\to\Pgth$ misidentification
probability amounts to $11\,(3)\%$ in the $\etau$ ($\tautau$) final state. The uncertainty in the $\Pgm
\to\Pgth$ misidentification probability is $12\,(5)\%$ in the $\mutau$ ($\tautau$) final state.
\item
The uncertainty in the $\Pgth$ identification efficiency is found to be $5\%$ per $\Pgth$ candidate.
It is factorized into a $4\,(8)\%$ part that is correlated and a $3\,(6)\%$ part that is uncorrelated
across all final states containing hadronic $\tau$ lepton decays in the $\etau$ and $\mutau$
($\tautau$) final states. A 7\% uncertainty in the $\Pgth$ trigger efficiency measurement is added
to the uncorrelated part in the $\tautau$ final state. The uncertainties related to the $\Pgth$
reconstruction and identification are applied to all processes that have been estimated from simulation
and that contain genuine hadronic $\tau$ lepton decays.
\item
The uncertainty in the jet energy scale affects the number of events entering each category. It is
applied to all processes estimated from simulation and ranges from $1$ to $6\%$, depending on the
final state and subcategory. Similarly, uncertainties in the rate with which both light-flavor jets
and genuine b jets pass the b tagging discriminator selection are applied to all processes estimated
from simulation. These uncertainties range from 1 to 5\%.
\item
Uncertainties in the resolution and response of the $\ptmiss$ are derived as part of the determination
of the recoil corrections. This leads to uncertainties ranging from 1 to 5\% that are incorporated
for all processes estimated from simulation and to which recoil corrections are applied. These are
all signal processes, Drell--Yan production and $\Wjets$ events. For the single $\cPqt$ quark, diboson and
$\ttbar$ backgrounds, which do not have recoil corrections applied, jet energy scale and unclustered
energy scale variations are propagated to the $\ptmiss$, also leading to uncertainties ranging
from 1 to 5\%.
\item
The uncertainty in the background yield from single $\cPqt$ quark and diboson production amounts to 5\%,
based on CMS measurements~\cite{Sirunyan:2016cdg,Khachatryan:2016tgp}. In the $\emu$ final state,
where the $\Wjets$ contribution is taken from simulation, the theoretical uncertainty in the cross
section calculation is 4\%. Due to the inclusion of the $\PZ\to\mumu$ and the $\ttbar$ control regions
in the model for the statistical inference of the signal, which control the Drell--Yan and $\ttbar$
normalization respectively, no theoretical cross section uncertainties are applied for these processes.
However, uncertainties are applied to the $\PZ\to\tau\tau$, $\PZ\to\ell\ell$, and $\ttbar$ processes
in all signal categories to account for the extrapolation from the control region to the signal regions.
The $\PZ\to\tau\tau$ extrapolation uncertainties range from 1 to 7\%. The extrapolation uncertainties
for $\PZ\to\ell\ell$ events are 4\%. The extrapolation uncertainties from the $\ttbar$ control region
to the signal regions are found to be below $1\%$. An additional uncertainty of 1\% is, however,
applied in the $\ttbar$ control region to account for fluctuations in the variables used to select
the events in this control region. The uncertainty in the estimation of the backgrounds in the DR$_{
i}$, which are taken from the simulation and subtracted from the data, for the determination of the
$\FF^{i}$ amounts to $3\,(4)\%$ in the $\etau$ and $\mutau$ ($\tautau$) final states.
\item
Since the background from QCD multijet events in the $\emu$ final state is determined from a control
region, uncertainties that account for the statistical uncertainty in the data and the subtracted
backgrounds in this control region are applied. In addition, this background is subject to uncertainties
related to the extrapolation from the control region to the signal regions. An overall 30\%
extrapolation uncertainty is applied, in addition to category-dependent uncertainties ranging from
4 to 29\%, in the measurement of the OS to SS transfer factor.
\item
Theoretical uncertainties in the acceptance of signal events in the associated production with b
quarks are obtained from variations of the renormalization ($\mu_{\text{r}}$) and factorization ($
\mu_{\text{f}}$) scales and of the internal ge\-ne\-ra\-tor matching scale $Q_{\text{sh}}$ related
to parton showering. The scales $\mu_{\text{r}}$ and $\mu_{\text{f}}$ are varied by factors of 0.5
and 2. The scale uncertainty is obtained from the envelope of the six variations of $\mu_{\text{r}}$
and $\mu_{\text{f}}$, as recommended in Ref.~\cite{deFlorian:2016spz}. Depending on the tested mass
it ranges between $-4\%$ (for 90\GeV), $-0.4\%$ (for 500\GeV), and $-2.5\%$ (for 3.2\TeV) in the
\text{b-tag} categories, and 0.8\% (for 90\GeV), 0.3\% (for 500\GeV), and 2.0\% (for 3.2\TeV) in the
\text{no b-tag} categories. The scale $Q_{\text{sh}}$ is varied by factors of $1/\sqrt{2}$ and $\sqrt{
2}$. The resulting uncertainty ranges between $-13.2\%$ (for 90\GeV), $-4.6\%$ (for 500\GeV), and
$-1.8\%$ (for 3.2\TeV) in the \text{b-tag} categories, and 2.6\% (for 90\GeV), 2.9\% (for 500\GeV),
and 1.4\% (for 3.2\TeV) in the \text{no b-tag} categories. The uncertainty from the variation of
$\mu_{\text{r}}$ and $\mu_{\text{f}}$, and the uncertainty from the variation of $Q_{\text{sh}}$
are added linearly, following the recommendation in Ref.~\cite{deFlorian:2016spz}.
\item
For the parameter scan in the model interpretations, theoretical uncertainties due to the different
choices of the factorization and renormalization scales in the signal predictions are included. The
MSTW2008~\cite{Martin:2009iq} PDFs are used for the calculation of the production cross sections. The
uncertainties in the choice for the PDFs are calculated following the recommended prescription given
in Refs.~\cite{Martin:2009iq,Martin:2009bu}. The uncertainties are evaluated separately for each
$\mA$--$\tan\beta$ point. They vary between 15 and 25\%.
\item
For all results shown in the following the SM Higgs boson production is taken into account in the
likelihood ratio. Uncertainties due to different choices of the renormalization and factorization
scales for the calculation of the production cross section of the SM Higgs boson amount to 3.9\%
for gluon fusion, 0.4\% for VBF, 2.8\% for $\PZ\PH$, and 0.5\% for $\PW\PH$ production. Uncertainties
due to different choices for the PDFs and $\alpha_{\text{s}}$ amount to 3.2\% for gluon fusion,
2.1\% for VBF, 1.6\% for $\PZ\PH$, and 1.9\% for $\PW\PH$ production. The procedure for deriving
these uncertainties is further described in Ref.~\cite{deFlorian:2016spz}.
\end{itemize}

The following systematic uncertainties allow correlated changes across bins that alter the shape
of the $\mTtot$ input distributions, and are referred to as shape uncertainties hereafter:

\begin{itemize}
\item
In the $\emu$ final state, shape uncertainties are applied to all processes with jets misidentified
as electrons or muons to account for the uncertainties in the jet$\to\Pe$ and jet$\to\Pgm$
misidentification probability. The size of these uncertainties depends on the jet $\pt$, with a
minimum uncertainty of $13\,(10)\%$ for electrons (muons).
\item
Three independent uncertainties are applied on the energy scale for genuine $\tau$ leptons decaying
hadronically; for the decay into a single charged hadron with and without neutral pions and the decay
into three charged hadrons. Each uncertainty is 1.2\%. They affect both the normalization and the shape
of the $\mTtot$ distribution for the signal, the $\PZ\to\tau\tau$, $\ttbar$ and diboson backgrounds
containing genuine $\tau$ leptons in the $\etau$, $\mutau$, and $\tautau$ final states.
\item
An asymmetric uncertainty of $+5\%\times\pt[\unit{TeV}]$ and $-35\%\times\pt[\text{TeV}]$ is applied
to account for the extrapolation in the $\Pgth$ identification efficiency estimate, which is mostly
determined by low-$\pt$ hadronic $\tau$ lepton decays close to the $\PZ$ boson peak, to higher-$\pt$
regimes of the $\tau$ leptons that are particularly relevant for the high-mass signal hypotheses. The
$\pt$ of the $\Pgth$ candidate is scaled by the corresponding factor. This uncertainty is applied to
the signal, the $\PZ\to\tau\tau$, $\ttbar$, and diboson backgrounds containing genuine $\tau$ leptons
in the $\etau$, $\mutau$, and $\tautau$ final states.
\item
In the $\etau$ final state, an uncertainty in the energy scale of electrons misidentified as hadronic
$\tau$ lepton decays is applied, split into a $1\,(0.5)\%$ uncertainty in the correction for the decay
mode with one charged hadron with (without) neutral pions. This uncertainty is only applied to the
$\PZ\to\ee$ process where one of the electrons is misidentified as a hadronic $\tau$ lepton decay.
\item
In the $\emu$ final state, an uncertainty in the electron energy scale is applied that amounts to 1\%
in the barrel and 2.5\% in the endcaps. In the $\etau$ final state this uncertainty is covered by the
uncertainty in the energy scale of the $\Pgth$ candidate.
\item
An uncertainty in the correction of the \pt of the top quarks in simulated $\ttbar$ events is
applied that corresponds to 100\% of the correction as discussed in Section~\ref{subsec:MC-backgrounds}.
It affects this background in all signal regions and in the $\ttbar$ control region. It is further
constrained by the $\ttbar$ control region described in Section~\ref{sec:event-selection}.
\item
Five uncertainties are included to cover the uncertainty in the reweighting method used to improve
the simulation of Drell--Yan events as described in Section~\ref{subsec:MC-backgrounds}. These
uncertainties include the propagation of the 0.2\% muon energy scale uncertainty to the derived
weights and the propagation of a 6\% $\ttbar$ cross section uncertainty, which affects the simulated
$\ttbar$ background that needs to be subtracted in the $\PZ\to\mumu$ selection. Since the reweighting
is obtained prior to the statistical inference for the signal this is not coupled to the $\ttbar$
control region. In addition, the statistical uncertainties in the measured weights are found to be
nonnegligible in three of the bins used to derive the correction, which leads to three additional
shape uncertainties related to the reweighting procedure.
\end{itemize}

In the $\mutau$, $\etau$, and $\tautau$ final states, the following shape uncertainties related
to the fake factor method are applied to those background components that are estimated by this method:

\begin{itemize}
\item
Statistical uncertainties in the estimate of the $\FF^{i}$ in the DR$_{i}$ are obtained from the
uncertainties of the fit used to parametrize the $\FF^{i}$. They amount to 4\% in the $\mutau$ final
state and range between 4 and 7\% (2 and 3\%) in the $\etau$ ($\tautau$) final states.
\item
In the $\etau$ and $\mutau$ final states, uncertainties are taken into account in the corrections due
to the finite number of events or omitted dependencies during the determination of the $\FF^{i}$.
This is done for all backgrounds considered. Additional uncertainties are taken into account in all
process specific corrections that are applied to the $\FF^{i}$. For $\FF^{\text{QCD}}$ these are the
correction of the extrapolation from the SS to the OS region and the correction as a function of the
lepton isolation. For $\FF^{\Wjets}$ this is the correction as a function of $\mTem$. For $\FF^{\ttbar
}$ this is the data-to-simulation correction in the dedicated control region. All these uncertainties
are added in quadrature for each corresponding background and vary between 7 and 10\% and between 5
and 7\% in the $\etau$ and $\mutau$ final states respectively.
\item
In the $\tautau$ final state, uncertainties are taken into account in the corrections due to the finite
number of events or omitted dependencies during the determination of the $\FF^{i}$. Additional uncertainties
in the correction of the SS to OS extrapolation as a function of the \pt of the other $\Pgth$
candidate, in the estimate of the fractions of $\Wjets$, Drell--Yan, and $\ttbar$ events with one jet
misidentified as a hadronic $\tau$ lepton decay, and in the use of $\FF^{\text QCD}$ for the estimation
of the $\Wjets$ and $\ttbar$ contributions to the total $\text{jet}\to\Pgth$ background are taken into
account. When added in quadrature, these additional uncertainties are of the order of 10\%.
\end{itemize}

The shape uncertainties related to the fake factor method are factorized into a pure shape and pure
normalization part. The normalization terms of the statistical uncertainties are added in quadrature
for each individual category in each final state and applied as normalization uncertainties.

In addition, uncertainties due to the limited population of the template distributions used for the
prediction of the background processes are taken into account by allowing each bin of each background
template to vary within its statistical uncertainty. These uncertainties are uncorrelated across the
bins of the input distributions. An overview of all uncertainties that have been taken into account
in the likelihood model used for the statistical analysis is given in Table~\ref{tab:systematics}.

\section{Results}
\label{sec:results}

The complete model, to extract the signal, results in a likelihood function of the form
\begin{linenomath}
  \begin{equation}
    \mathcal{L}\left(\left.\{k_{i}\}\right|\mu s(\theta)+b(\theta)\right) =
    \prod\limits_{i}\mathcal{P}(k_{i}|\mu s_{i}(\theta)+b_{i}(\theta))
    \,
    \prod\limits_{j}\mathcal{C}(\hat{\theta}_{j}|\theta_{j}),
    \label{eq:likelihood}
  \end{equation}
\end{linenomath}
where $i$ labels all bins of the input distributions with event numbers $k_{i}$ in all event
subcategories and control regions and $j$ all nuisance parameters, referred to by $\theta$. The term
$\theta_{j}$ corresponds to a given nuisance parameter, $\mu$ to a scaling parameter for a given signal
$s_{i}$, and $b_{i}$ to the prediction of all backgrounds in bin $i$. The function $\mathcal{P}(k_{i}
|\mu s_{i}(\theta)+b_{i}(\theta))$ corresponds to a Poisson distribution, $\mathcal{C}(\hat{\theta}_{
j}|\theta_{j})$ to the probability density function used to implement the uncertainty related to the
nuisance parameter $\theta_{j}$, and $\hat{\theta}_{j}$ to the estimate for $\theta_{j}$ from the fit
to the data. All distributions shown in Figs.~\ref{fig:mTtot-distributions-1} and~\ref{fig:mTtot-distributions-2}
are after an MSSM $\mhmodp$ signal-plus-background hypothesis, corresponding to $\mA=700\GeV$ and
$\tan\beta=20$, has been fitted to the data. No signal is observed in the investigated mass range
between 90\GeV and 3.2\TeV and upper limits on the presence of a signal are set in the two
interpretations of the data as discussed in Section~\ref{sec:signal-extraction}. This is done following
the modified frequentist approach as described in Refs.~\cite{Junk:1999kv,Read:2002hq}, using the same
definition of the test statistic as in the search for the SM Higgs
boson~\cite{CMS-NOTE-2011-005,Chatrchyan:2012tx}:
\begin{linenomath}
  \begin{equation}
    q_{\mu} = -2\ln
    \left(
      \frac{\mathcal{L}(\left.\{k_{i}\}\right|\mu s(\hat{\theta}_{\mu})+b(\hat{\theta}_{\mu}))}
      {\mathcal{L}(\left.\{k_{i}\}\right|\hat{\mu} s(\hat{\theta}_{\hat{\mu}})+b(\hat{\theta}_{\hat{\mu}}))}
    \right) , \;\; 0\leq \hat{\mu} \leq \mu,
    \label{eq:likelihood-ratio}
  \end{equation}
\end{linenomath}
where the hat in $\hat{\mu}$, $\hat{\theta}_{\mu}$ and $\hat{\theta}_{\hat{\mu}}$ again indicates the
estimate of the corresponding quantity from the fit to the data and the index of $q_{\mu}$ indicates
that the fit to the data has been performed for a fixed value of $\mu$. In the large number limit the
distribution of $q_{\mu}$ can be approximated by analytic functions, from which the median and the
uncertainty contours can be obtained as described in Ref.~\cite{Cowan:2010js}.

\begin{table}[htbp]
  \topcaption{
    Overview of the systematic uncertainties used in the likelihood model for the statistical inference
    of the signal. The label ``MC'' refers to all processes that are obtained from simulation, the label
    ``\FF'' refers to all backgrounds that are obtained from the fake factor method. Values in parentheses
    correspond to additional uncertainties correlated across final states or event categories. Detailed
    descriptions are given in Section~\ref{sec:uncertainties}.
  }
  \label{tab:systematics}
  \centering
  \begin{tabular}{lccccccc}
    Uncertainty & $\emu$ & $\etau$ & $\mutau$ & $\tautau$ & Process & Shape & Variation \\
    \hline
    Integrated luminosity & $\checkmark$ & $\checkmark$ & $\checkmark$ & $\checkmark$ & MC & \NA & 2.5\% \\
    $\text{Jet}\to\Pe$ mis-ID & $\checkmark$ & \NA & \NA & \NA & MC & $\checkmark$ & 13\% \\
    $\text{Jet}\to\mu$ mis-ID & $\checkmark$ & \NA & \NA & \NA & MC & $\checkmark$ & 10\% \\
    \multirow{2}{*}{e/$\mu$-trigger, ID, isolation} & $\checkmark$ & $\checkmark$ & \NA & \NA & MC
    & \NA & 2\% \\
    & $\checkmark$ & \NA & $\checkmark$ & \NA & MC & \NA & 2\% \\
    \multirow{2}{*}{$\Pe\to\Pgth$ mis-ID} & \NA & $\checkmark$ & \NA & \NA & $\PZ\to\ee$ & \NA & 11\% \\
    & \NA & \NA & \NA & $\checkmark$ & $\PZ\to\ee$ & \NA & 3\% \\
    \multirow{2}{*}{$\mu\to\Pgth$ mis-ID} & \NA & \NA & $\checkmark$ & \NA & $\PZ\to\mumu$ & \NA & 12\% \\
    & \NA & \NA & \NA & $\checkmark$ & $\PZ\to\mumu$ & \NA & 5\% \\
    $\Pgth$-trigger & \NA & \NA & \NA & $\checkmark$ & MC & \NA & 7\% \\
    \multirow{2}{*}{$\Pgth$-ID} & \NA & $\checkmark$ & $\checkmark$ & \NA & MC & \NA & $3\,(4)\%$ \\
    & \NA & \NA & \NA & $\checkmark$ & MC & \NA & $6\,(8)\%$ \\
    $\Pgth$-ID (high $\pt$) & \NA & $\checkmark$ & $\checkmark$ & $\checkmark$ & MC & $\checkmark$ & \pt dep. \\
    $\Pgth$ energy scale & \NA & $\checkmark$ & $\checkmark$ & $\checkmark$ & MC & $\checkmark$ & 1.2\% \\
    $\Pe\to\Pgth$ energy scale & \NA & $\checkmark$ & \NA & \NA & $\PZ\to\ee$ & $\checkmark$ & 0.5--1.0\% \\
    $\Pe$ energy scale & $\checkmark$ & \NA & \NA & \NA & MC & $\checkmark$ & 1.0--2.5\% \\
    Jet energy scale & $\checkmark$ & $\checkmark$ & $\checkmark$ & $\checkmark$ & MC & \NA & 1--6\% \\
    b tagging & $\checkmark$ & $\checkmark$ & $\checkmark$ & $\checkmark$ & MC & \NA & 1--5\% \\
    $\ptmiss$ resp./res. & $\checkmark$ & $\checkmark$ & $\checkmark$ & $\checkmark$ & MC & \NA & 1--5\% \\
    \multirow{3}{*}{Bkgr. in signal categories} & $\checkmark$ & $\checkmark$ & $\checkmark$ & $\checkmark$ & Diboson & \NA & 5\% \\
    & $\checkmark$ & $\checkmark$ & $\checkmark$ & $\checkmark$ & single $\cPqt$ & \NA & 5\% \\
    & $\checkmark$ & \NA & \NA & \NA & $\Wjets$ & \NA & $4\%$ \\
    \multirow{4}{*}{Sideband extrapolation} & $\checkmark$ & $\checkmark$ & $\checkmark$ & $\checkmark$ & $\PZ\to\tau\tau$ & \NA & 1--7\% \\
    & $\checkmark$ & $\checkmark$ & $\checkmark$ & $\checkmark$ & $\PZ\to\ell\ell$ & \NA & 4\% \\
    & $\checkmark$ & $\checkmark$ & $\checkmark$ & $\checkmark$ & $\ttbar$ & \NA & 1\% \\
    & $\checkmark$ & \NA & \NA & \NA & QCD & \NA & 4--$29\,(30)\%$ \\
    Top quark \pt reweighting & $\checkmark$ & $\checkmark$ & $\checkmark$ & $\checkmark$ & $\ttbar$ & $\checkmark$ & 100\% \\
    $\PZ$ reweighting of LO MC & $\checkmark$ & $\checkmark$ & $\checkmark$ & $\checkmark$ & $\PZ\to\tau\tau, \ell\ell$ & $\checkmark$ & See text \\
    \multirow{2}{*}{Bkgr. in DR$_{\text QCD/\Wjets}$} & \NA & $\checkmark$ & $\checkmark$ & \NA & MC & \NA & 3\% \\
    & \NA & \NA & \NA & $\checkmark$ & MC & \NA & 4\% \\
    \multirow{3}{*}{$\FF^{i}$ stat. uncert.} & \NA & $\checkmark$ & \NA & \NA & \FF & $\checkmark$ & 4--7\% \\
    & \NA & \NA & $\checkmark$ & \NA & \FF & $\checkmark$ & 4\% \\
    & \NA & \NA & \NA & $\checkmark$ & \FF & $\checkmark$ & 2--3\% \\
    \multirow{3}{*}{$\FF^{i}$ corrections} & \NA & $\checkmark$ & \NA & \NA & \FF & $\checkmark$ & 7--10\% \\
    & \NA & \NA & $\checkmark$ & \NA & \FF & $\checkmark$ & 5--7\% \\
    & \NA & \NA & \NA & $\checkmark$ & \FF & $\checkmark$ & 10\% \\
    b-associated signal acceptance & $\checkmark$ & $\checkmark$ & $\checkmark$ & $\checkmark$ & Signal & \NA & 3.2--16.5\% \\
    \multirow{2}{*}{PDF/scale} & $\checkmark$ & $\checkmark$ & $\checkmark$ & $\checkmark$ & Signal & \NA & 15--25\% \\
    & $\checkmark$ & $\checkmark$ & $\checkmark$ & $\checkmark$ & SM Higgs & \NA & 0.5--3.2\% \\
    \hline
  \end{tabular}
\end{table}

\begin{figure}[h!]
  \centering
  \includegraphics[width=0.48\textwidth]{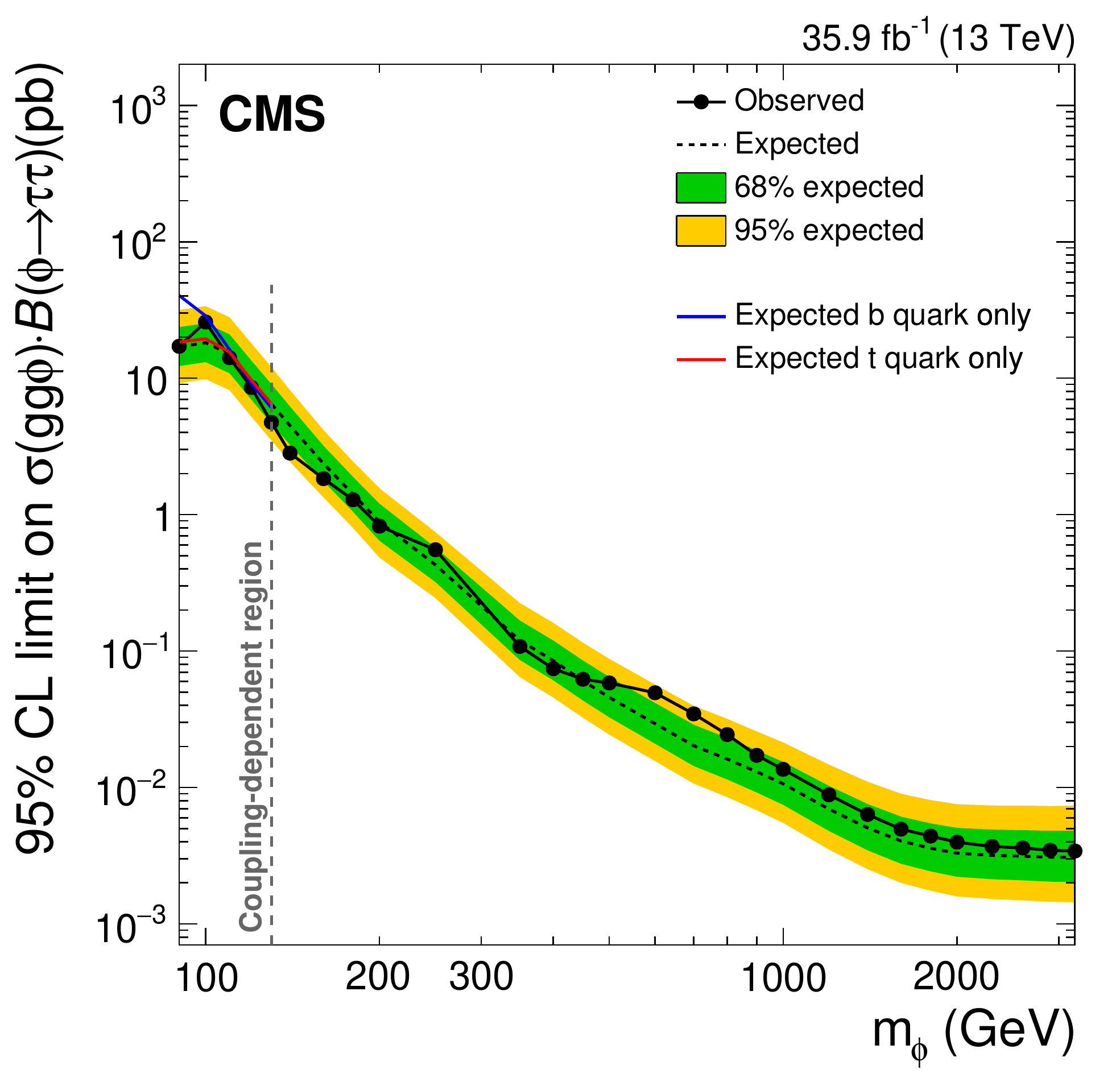}
  \includegraphics[width=0.48\textwidth]{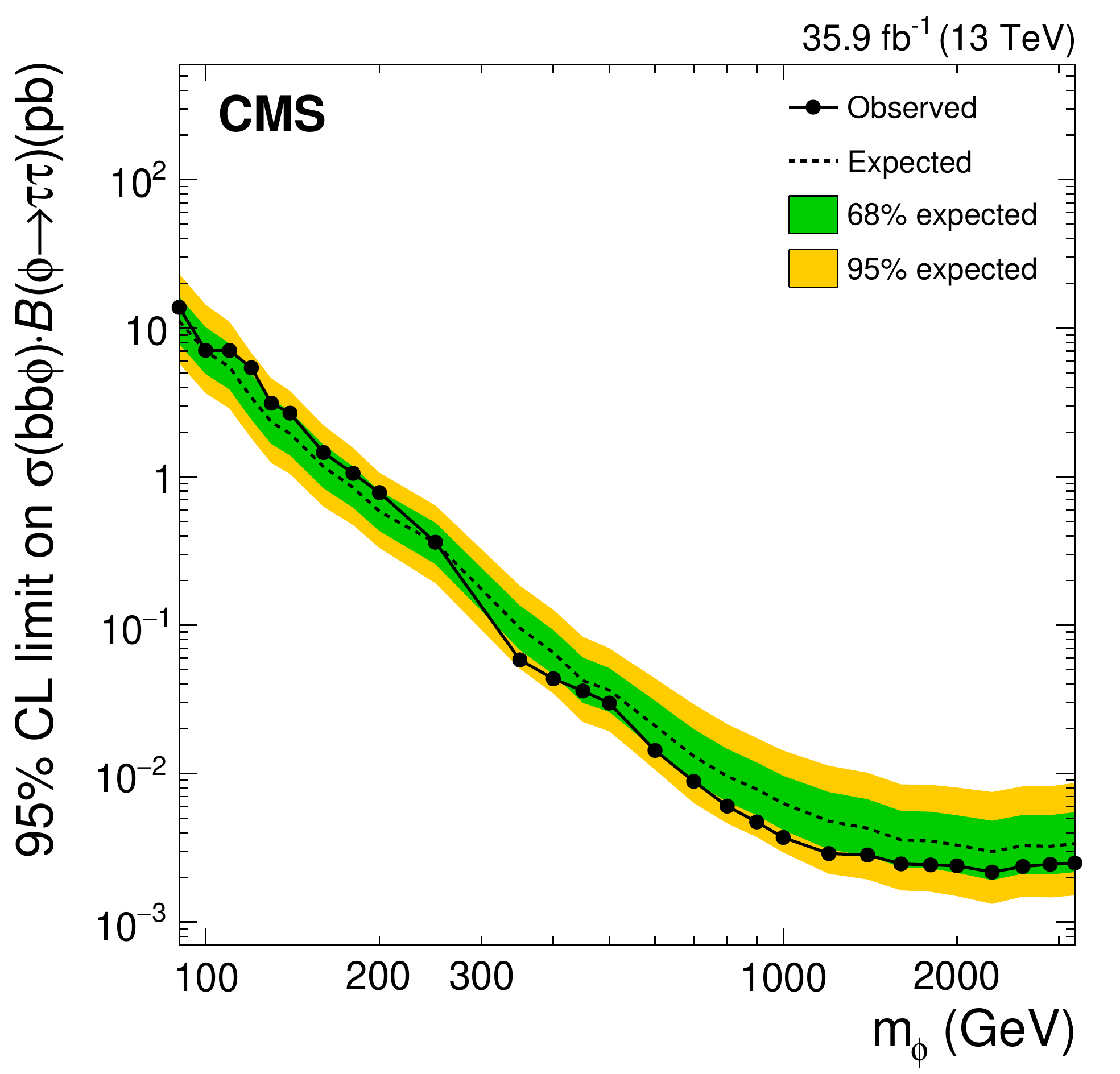}
  \caption{
    Expected and observed 95\% \CL upper limits for the production of a single narrow resonance,
    $\phi$, with a mass between 90\GeV and 3.2\TeV in the $\tau\tau$ final state (left)
    for the production via gluon fusion (gg$\phi$) and (right) in association with b quarks (bb$\phi$).
    The expected median of the exclusion limit is shown by the dashed line. The dark green and bright
    yellow bands indicate the 68 and 95\% confidence intervals for the variation of the expected
    exclusion limit. The black dots correspond to the observed limits. In the left panel the expected
    exclusion limits for the cases where (blue continuous line) only the b quark and (red continuous
    line) only the t quark are taken into account in the fermion loop are also shown. Left of
    the dashed vertical line the two different assumptions lead to visible differences in the expected
    exclusion limit.
  }
  \label{fig:results_modelindep_cmb}
\end{figure}

\begin{figure}[htbp]
  \centering
  \includegraphics[width=0.45\textwidth]{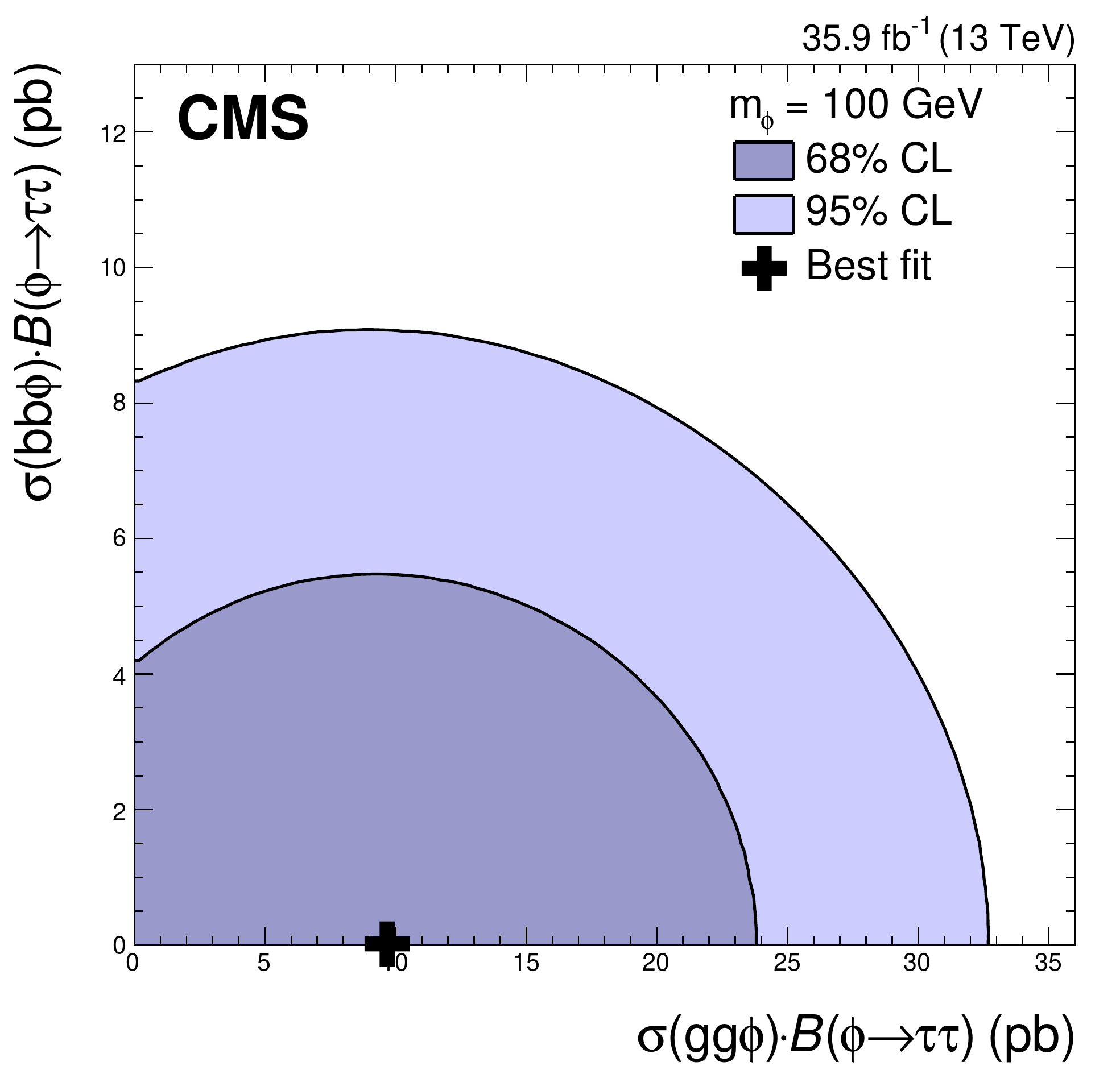}
  \includegraphics[width=0.45\textwidth]{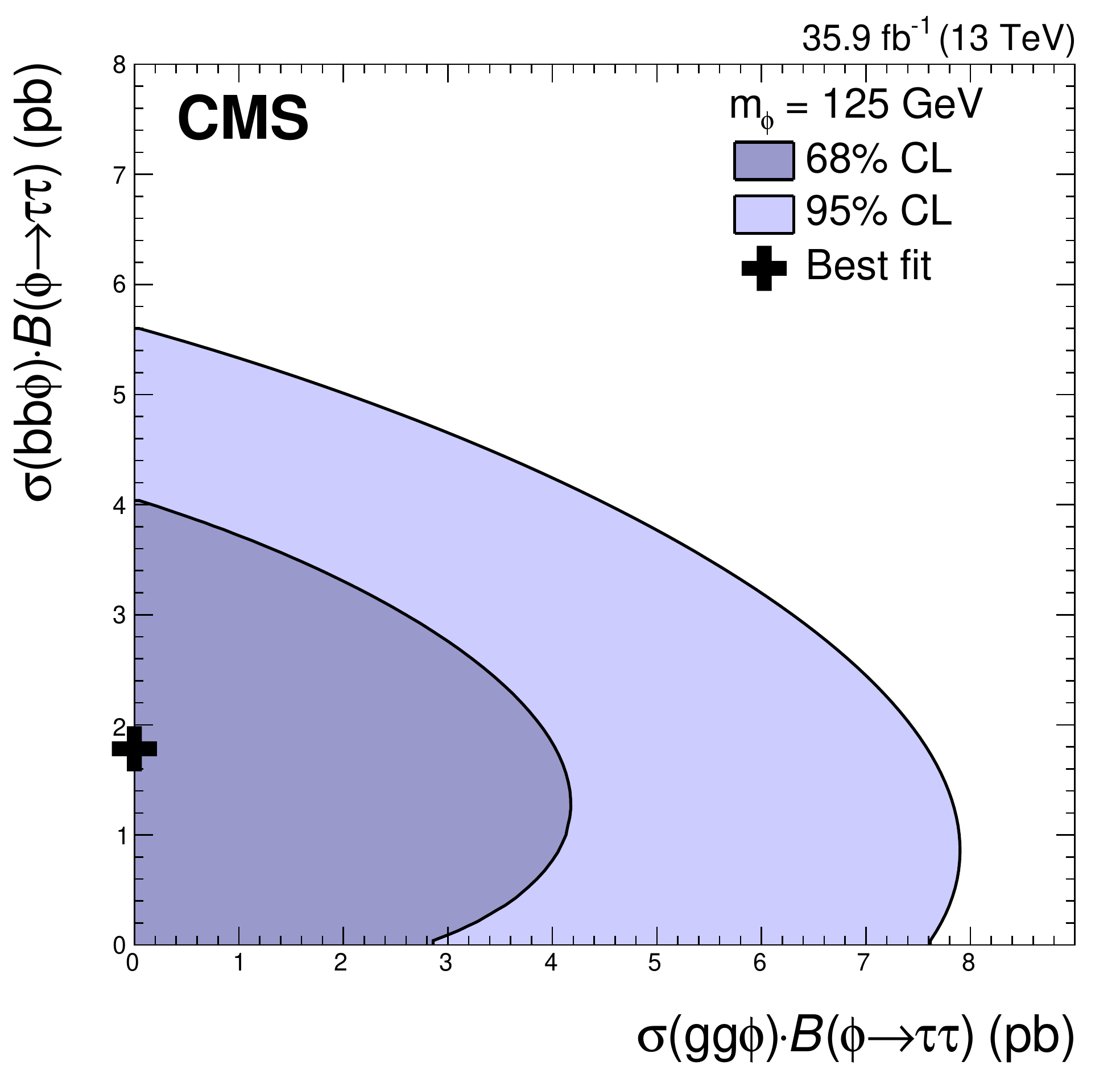}
  \includegraphics[width=0.45\textwidth]{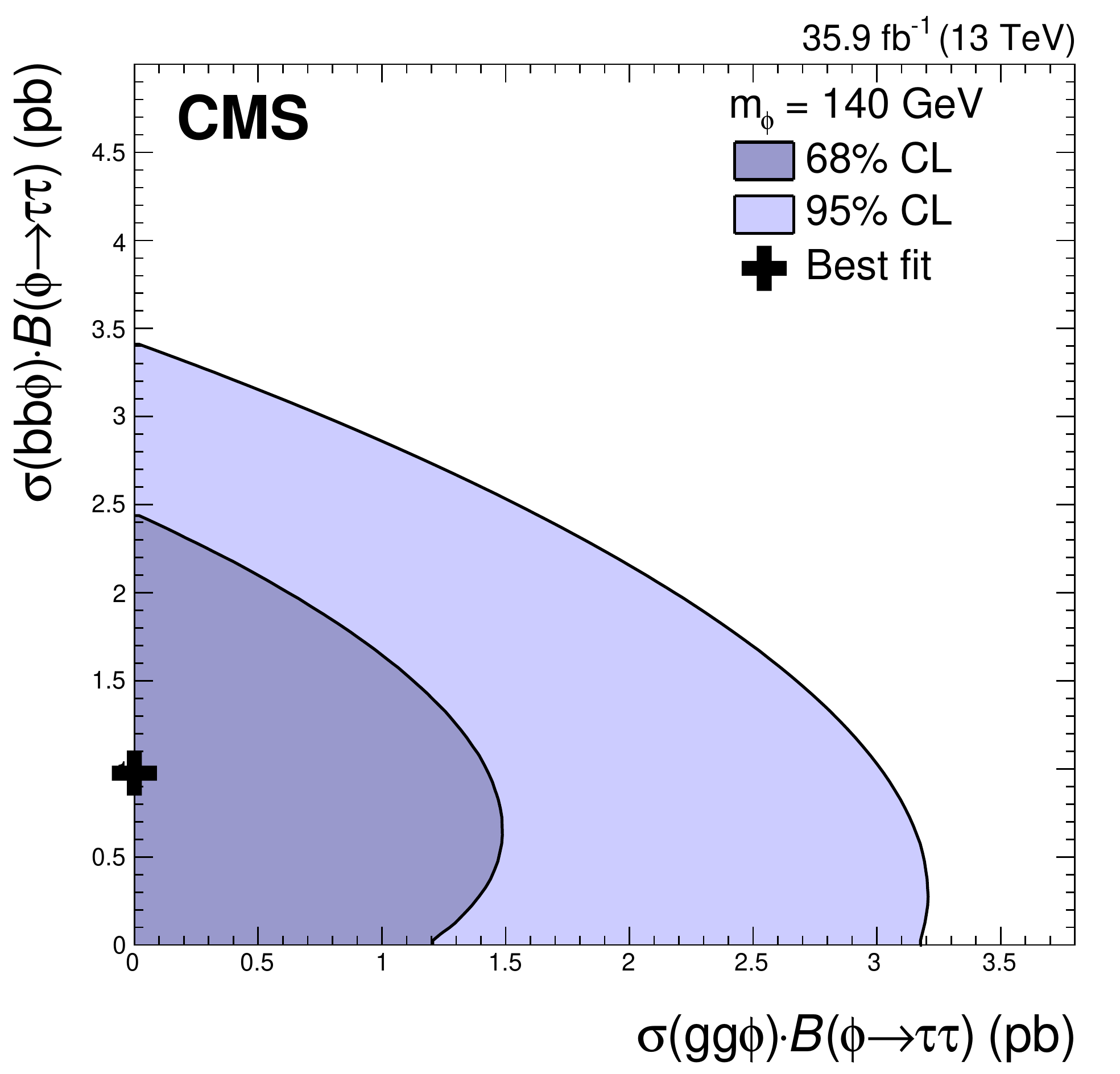}
  \includegraphics[width=0.45\textwidth]{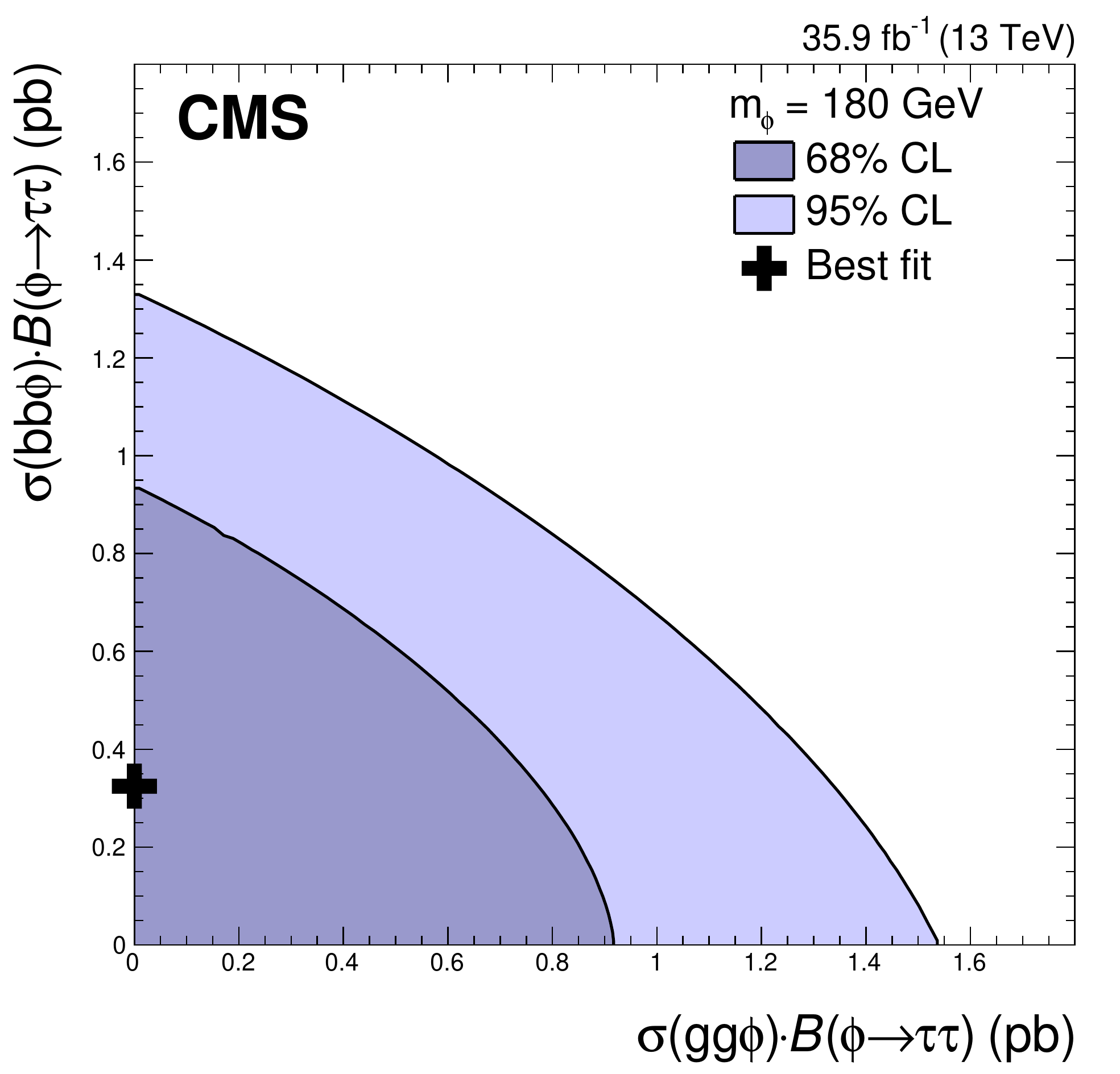}
  \includegraphics[width=0.45\textwidth]{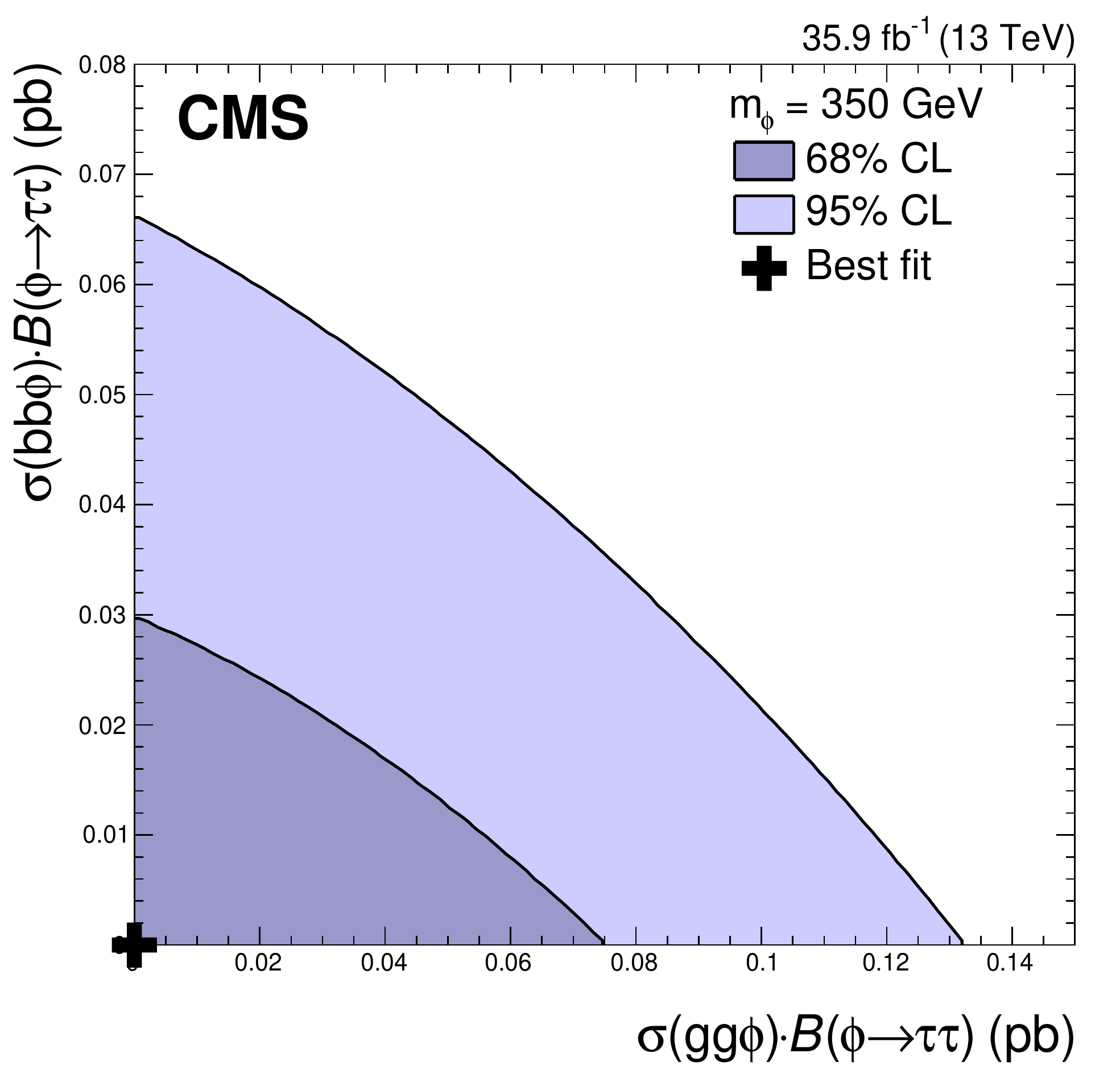}
  \includegraphics[width=0.45\textwidth]{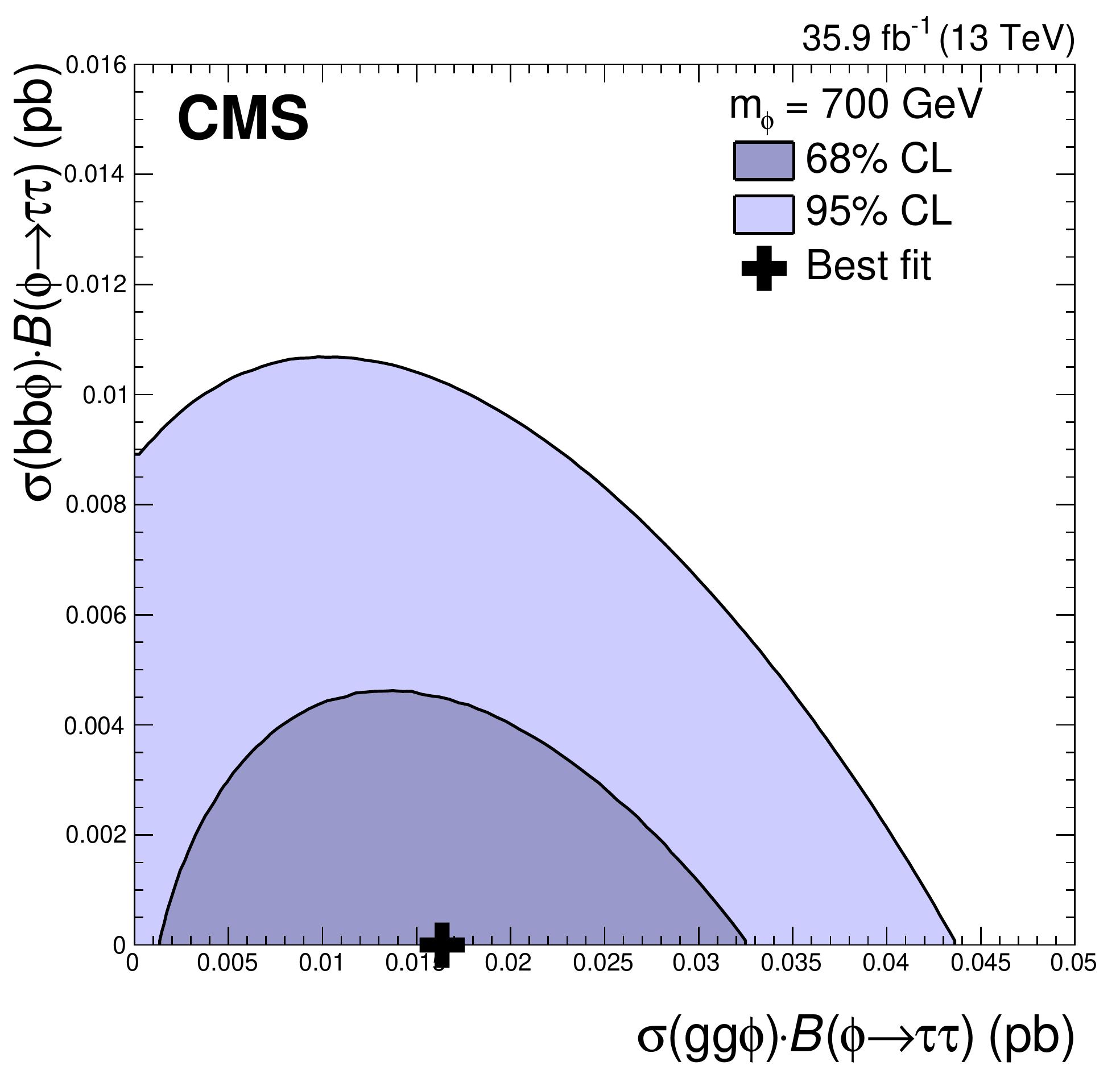}
  \caption{
    Scan of the likelihood function for the search in the $\tau\tau$ final state for a single narrow
    resonance, $\phi$, produced via gluon fusion ($\Pg\Pg\phi$) or in association with b quarks ($\PQb\PQb\phi$).
    A representative subset of the mass points tested at (upper left) 100\GeV, (upper right) 125\GeV,
    (middle left) 140\GeV, (middle right) 180\GeV, (lower left) 350\GeV, and (lower right) 700\GeV is
    shown. Note that in the fits the signal strengths are not allowed to become negative.
  }
  \label{fig:likelihood-scan}
\end{figure}

In the first interpretation of the data 95\% confidence level (\CL) upper limits are set on the
product of the branching fraction for the decay into $\tau$ leptons and the cross section for the
production of a single narrow width resonance, $\phi$, via gluon fusion or in association with b
quarks. In Fig.~\ref{fig:results_modelindep_cmb} these limits are shown as a function of $m_{\phi}$.
For the determination of the limit on one process, \eg, gluon fusion, the normalization for the
corresponding other process, \eg, associated production with b quarks, is treated as a freely varying
parameter in the signal-plus-background fit that is performed prior to the limit calculation. The
expectation for an SM Higgs boson at 125\GeV is taken into account in the SM backgrounds. For both
production modes the \pt spectrum of the $\phi$ is estimated at NLO precision in $\alpha_{
\text{s}}$, as described in Section~\ref{subsec:simulation}. Differences in the sensitivity of the
analysis only occur at low masses, where the \pt of the $\phi$ significantly contributes to
the \pt of its decay products. In the figure this is emphasized by adding the median for the
expected limit using either only the b quark or only the t quark for the modeling of the $\phi$
$\pt$ spectrum. For the production via gluon fusion the expected limits range between 18\unit{pb} at
$m_{\phi}=90\GeV$ and 3.5\unit{fb} at $m_{\phi}=3.2\TeV$. For the production in association with b
quarks they range between 15\unit{pb} (at $m_{\phi}=90\GeV$) and 2.5\unit{fb} (at $m_{\phi}=3.2\TeV$).
In both cases, the excluded cross section falls with increasing mass, before becoming constant at
around 1\TeV. No significant deviation from the expectation is observed. When restricted to the
$\etau$, $\mutau$, or $\tautau$ final state, the results obtained from the cross-checks summarized
in Section~\ref{subsec:cross-checks} are compatible with the results obtained from the main analysis
described in this paper. A scan of the likelihood for this signal model is also performed, as a
function of the gluon fusion cross section and the cross section for the associated production with
b quarks, for the tested mass points. A representative subset of this likelihood scan at six mass
points is shown in Fig.~\ref{fig:likelihood-scan}.

In the second interpretation of the data, exclusion contours in the $\mA$--$\tan\beta$ plane
are determined for two representative benchmark scenarios of the MSSM, the $\mhmodp$
and the hMSSM~\cite{Maiani:2013hud,Djouadi:2013uqa,Djouadi:2015jea}. Apart from small phase space
regions, the $\mhmodp$ scenario is compatible with the observation of the Higgs boson at 125\GeV,
which is interpreted as the $\Ph$ within the theoretical uncertainties in $m_{\Ph}$ of ${\pm}3
\GeV$~\cite{Degrassi:2002fi,Allanach:2004rh}. The phenomenological hMSSM also incorporates the observed
Higgs boson with a fixed mass of $125\GeV$, interpreting it as the $\Ph$. The uncertainties in the
mass measurement are then used in turn to estimate the main radiative corrections to predict the
masses and couplings of the remaining MSSM Higgs bosons. For the determination of the exclusion
contours the model predictions as provided by the LHC Higgs Cross Section Working
Group~\cite{Heinemeyer:2013tqa,deFlorian:2016spz} are used. Inclusive cross sections for the production
via gluon fusion are calculated using the program \textsc{SusHi} (v1.4.1)~\cite{Harlander:2012pb},
including NLO QCD corrections in the context of the MSSM~\cite{Spira:1995rr,Harlander:2004tp,
Harlander:2005rq,Degrassi:2010eu,Degrassi:2011vq,Degrassi:2012vt}, as well as NNLO QCD corrections
for the top quark contribution to the fermion loop in the heavy top quark limit~\cite{Harlander:2002wh,
Anastasiou:2002yz,Ravindran:2003um,Harlander:2002vv,Anastasiou:2002wq}, and electroweak effects from light
quarks~\cite{Aglietti:2004nj,Bonciani:2010ms}. For associated production with b quarks four-flavor
scheme NLO QCD calculations~\cite{Dittmaier:2003ej,PhysRevD.69.074027} and five-flavor scheme NNLO
QCD calculations, as implemented in \textsc{SusHi} based on \textsc{bbh@nnlo}~\cite{PhysRevD.68.013001},
are combined using the Santander matching scheme~\cite{Harlander:2011aa}. The Higgs boson masses and
mixing, and the effective Yukawa couplings for the $\mhmodp$ scenario, are calculated
using the \textsc{FeynHiggs} 2.10.2~\cite{Heinemeyer:1998yj,Heinemeyer:1998np,Degrassi:2002fi,
Frank:2006yh,Hahn:2013ria} code. The branching fraction of the MSSM Higgs bosons to $\tau$ leptons
is calculated with \textsc{FeynHiggs} for the $\mhmodp$ scenario and using the program
\textsc{hdecay} 6.40~\cite{Djouadi:1997yw} for the hMSSM scenario.

\begin{figure}[h]
  \centering
  \includegraphics[width=0.48\textwidth]{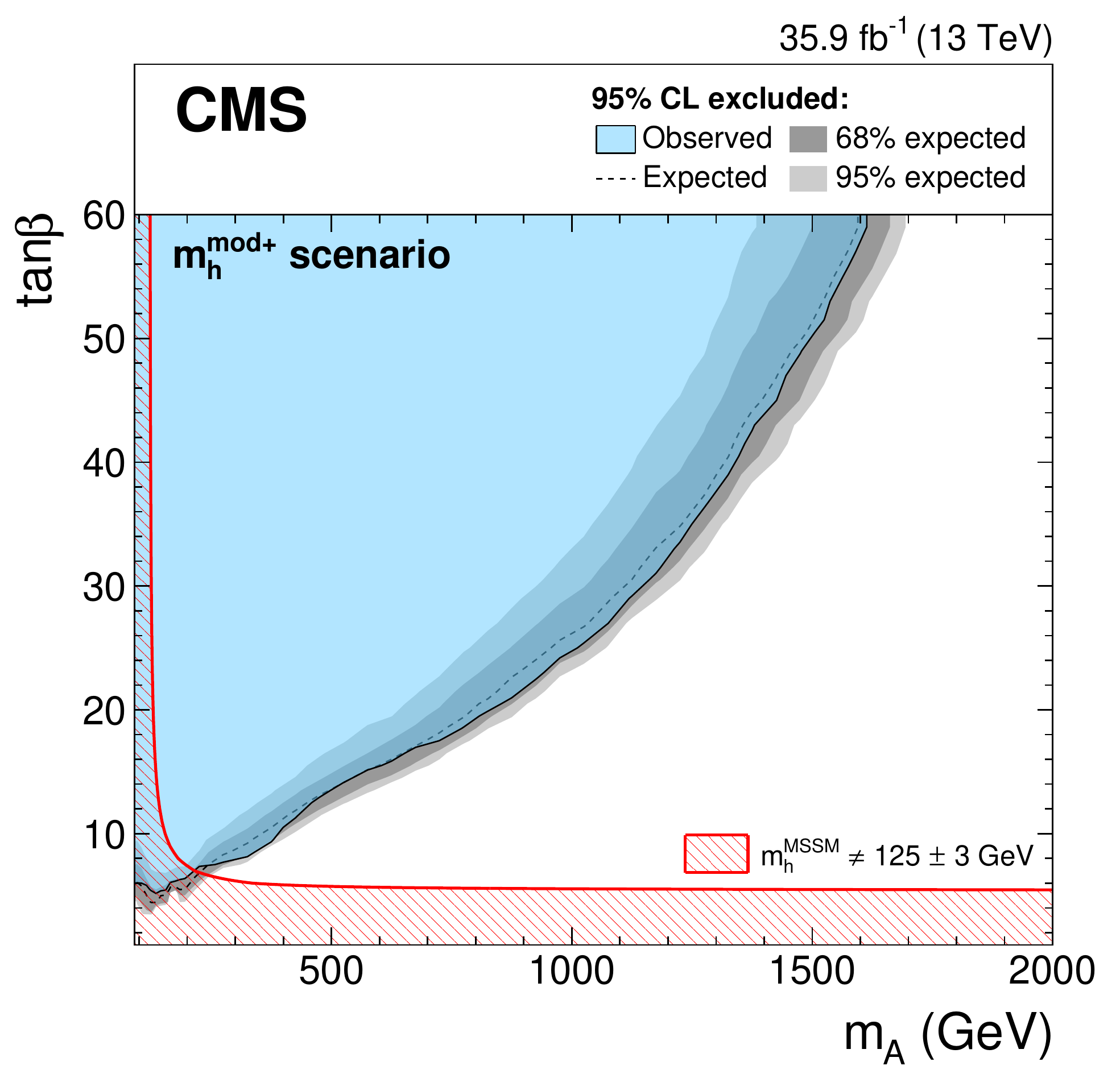}
  \includegraphics[width=0.48\textwidth]{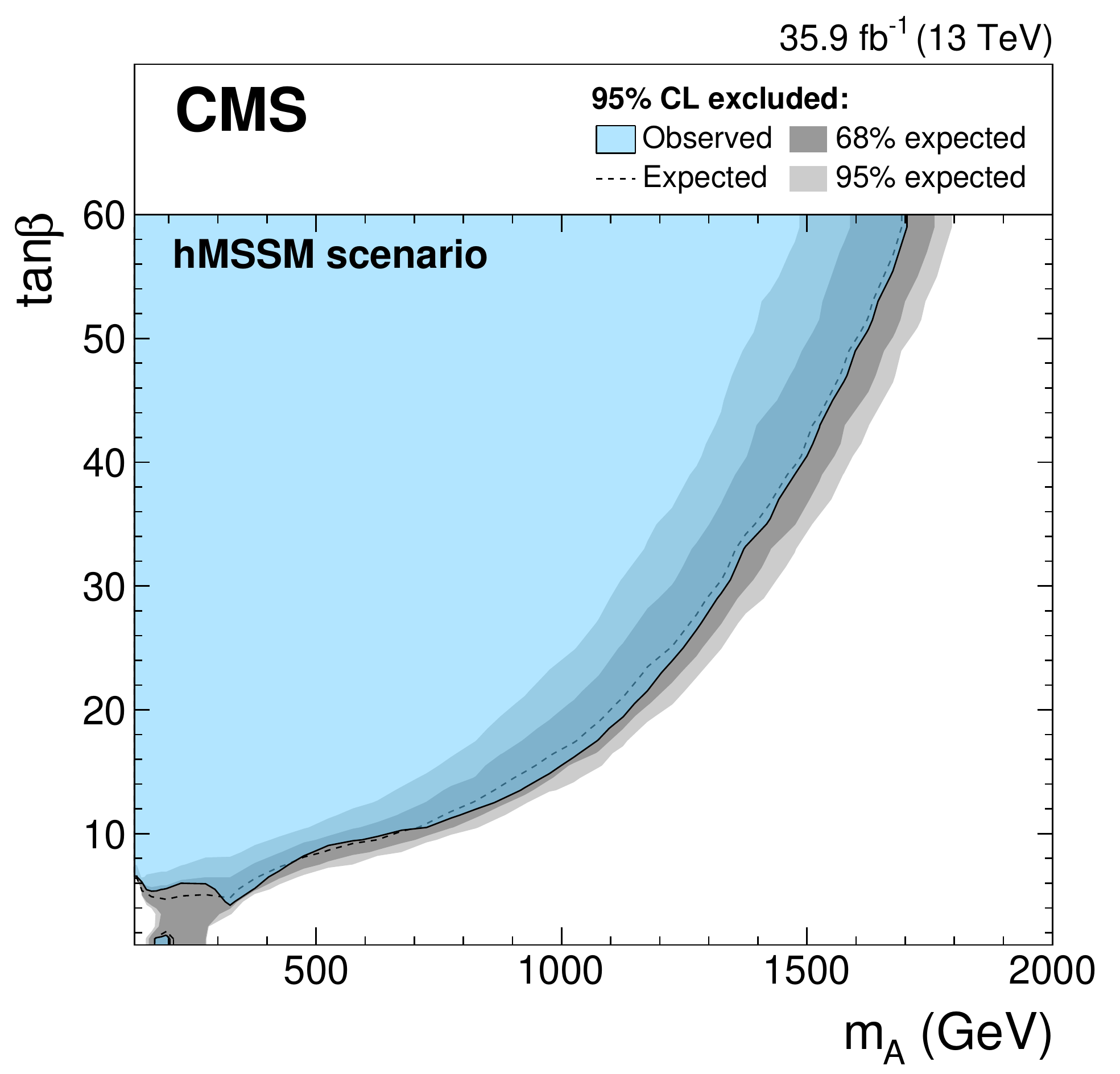}
  \caption{
    Expected and observed 95\% \CL exclusion contour (left) in the MSSM $\mhmodp$ and (right) in the
    hMSSM scenarios. The expected median is shown as a dashed black line. The dark and bright gray
    bands indicate the 68 and 95\% confidence intervals for the variation of the expected exclusion.
    The observed exclusion contour is indicated by the colored blue area. For the $\mhmodp$ scenario,
    those parts of the parameter space, where $m_{\Ph}$ deviates by more then ${\pm}3\GeV$ from the
    mass of the observed Higgs boson at 125\GeV are indicated by a red hatched area.
  }
  \label{fig:exclusion-contours}
\end{figure}

The simulated single neutral Higgs boson signals are combined into a multiresonance signal model for
the given values of $\mA$ and $\tan\beta$, taking into account the predictions for the mass,
production cross sections, and branching fraction into $\tau$ leptons for each of the neutral Higgs
bosons. For each value of $\mA$ and $\tan\beta$, using a fine-grain scan, a maximum likelihood
fit to the data is performed under the background-only and the signal-plus-background hypotheses using
the likelihood of Eq.~(\ref{eq:likelihood}) with a test statistic that is slightly different from
Eq.~(\ref{eq:likelihood-ratio}). The numerator remains the same, with a fixed value of $\mu=1$, and
corresponds to the signal prediction for the given value of $\mA$ and $\tan\beta$. However
no signal strength parameter is included in the denominator; the model is thus fixed to the
background-only prediction. Note that the SM Higgs boson is added to the non Higgs boson background
processes. This turns the likelihood ratio into a comparison between the MSSM and the SM Higgs sector
hypotheses, and ensures a well defined problem even when the analysis becomes sensitive to the observed
Higgs boson at 125\GeV. In such a situation a test of the MSSM hypothesis against a background
hypothesis ignoring the SM Higgs boson would be based on a wrong null-hypothesis. The median and
confidence intervals for the expected exclusion contour are determined from pseudo-experiments. In
Fig.~\ref{fig:exclusion-contours} the observed and expected 95\% \CL exclusion contours for the MSSM
$\mhmodp$ and the hMSSM scenarios are shown.
The exclusion contours reach up to 1.6\TeV, extending
the excluded mass range by almost a factor of two in $\mA$ compared to the previous CMS publication
using the same final state~\cite{Khachatryan:2014wca}. In both scenarios the exclusion contours extend
down to values of $\tan\beta\approx6$ for values of $\mA\lesssim 250\GeV$. For the $\mhmodp$ scenario,
those parts of the parameter space in which $m_{\Ph}$ deviates by more then ${\pm}3\GeV$ from the mass
of the observed Higgs boson at 125\GeV are indicated by a red hatched area. These results are compatible
with the findings of a similar search performed by the ATLAS collaboration, based on an equivalent
dataset~\cite{Aaboud:2017sjh}.

In the low mass region the exclusion contour is similar to the previous CMS
publication, while a higher sensitivity might be expected. This can be attributed to three main factors:
the choice of single lepton triggers in the $\etau$ and $\mutau$ final states together with the higher
instantaneous luminosity leads to the need for higher \pt thresholds at the trigger level and therefore
reduced signal acceptance; the change of the discriminating variable from the estimate of the fully
reconstructed $\tau\tau$ mass to $\mTtot$ provides more sensitivity for high masses, but slightly
less sensitivity for lower masses; and finally the prediction of the kinematic distributions of the
signal at NLO precision reveals a generally softer \pt spectrum for the gluon fusion production
mode, which dominates for low values of $\tan\beta$. Over the whole mass range the observed exclusion
contours follow the expectation with the largest deviations still contained in the 95\% confidence
interval for the variation of the expected exclusion.

\section{Summary}
\label{sec:summary}

A search for additional heavy neutral Higgs bosons in the decay into two $\tau$ leptons in the context
of the minimal supersymmetric standard model (MSSM) has been presented. This search has been performed
in the most sensitive $\emu$, $\etau$, $\mutau$, and $\tautau$ final states of the $\tau\tau$ pair,
where $\Pgth$ indicates a hadronic $\tau$ lepton decay. No signal has been found. Model-independent
limits at 95\% confidence level have been set for the production of a single narrow resonance decaying
into a pair of $\tau$ leptons. These range from 18\unit{pb} at 90\GeV to 3.5\unit{fb} at 3.2\TeV
for production via gluon fusion and from 15\unit{pb} (at 90\GeV) to 2.5\unit{fb} (at 3.2\TeV) for
production in association with b quarks. Finally 95\% confidence level exclusion contours have been
provided for two representative benchmark scenarios, namely the $\mhmodp$ and the hMSSM scenarios. In
these two scenarios the presence of a neutral heavy MSSM Higgs boson up to $m_{A}\lesssim250\GeV$ is
excluded for $\tan\beta$ values above 6. The exclusion contour reaches $1.6\TeV$ for $\tan\beta=60$.

\clearpage

\begin{acknowledgments}
We thank Emanuele Bagnaschi and Stefan Liebler for their support in developing the procedure for obtaining the Higgs boson kinematics at NLO plus parton shower precision with \POWHEG. This work has been performed in the frame of the LHC Higgs Cross Section Working Group.

We congratulate our colleagues in the CERN accelerator departments for the excellent performance of the LHC and thank the technical and administrative staffs at CERN and at other CMS institutes for their contributions to the success of the CMS effort. In addition, we gratefully acknowledge the computing centers and personnel of the Worldwide LHC Computing Grid for delivering so effectively the computing infrastructure essential to our analyses. Finally, we acknowledge the enduring support for the construction and operation of the LHC and the CMS detector provided by the following funding agencies: BMWFW and FWF (Austria); FNRS and FWO (Belgium); CNPq, CAPES, FAPERJ, and FAPESP (Brazil); MES (Bulgaria); CERN; CAS, MoST, and NSFC (China); COLCIENCIAS (Colombia); MSES and CSF (Croatia); RPF (Cyprus); SENESCYT (Ecuador); MoER, ERC IUT, and ERDF (Estonia); Academy of Finland, MEC, and HIP (Finland); CEA and CNRS/IN2P3 (France); BMBF, DFG, and HGF (Germany); GSRT (Greece); NKFIA (Hungary); DAE and DST (India); IPM (Iran); SFI (Ireland); INFN (Italy); MSIP and NRF (Republic of Korea); LAS (Lithuania); MOE and UM (Malaysia); BUAP, CINVESTAV, CONACYT, LNS, SEP, and UASLP-FAI (Mexico); MBIE (New Zealand); PAEC (Pakistan); MSHE and NSC (Poland); FCT (Portugal); JINR (Dubna); MON, RosAtom, RAS and RFBR (Russia); MESTD (Serbia); SEIDI, CPAN, PCTI and FEDER (Spain); Swiss Funding Agencies (Switzerland); MST (Taipei); ThEPCenter, IPST, STAR, and NSTDA (Thailand); TUBITAK and TAEK (Turkey); NASU and SFFR (Ukraine); STFC (United Kingdom); DOE and NSF (USA).

\hyphenation{Rachada-pisek} Individuals have received support from the Marie-Curie program and the European Research Council and Horizon 2020 Grant, contract No. 675440 (European Union); the Leventis Foundation; the A. P. Sloan Foundation; the Alexander von Humboldt Foundation; the Belgian Federal Science Policy Office; the Fonds pour la Formation \`a la Recherche dans l'Industrie et dans l'Agriculture (FRIA-Belgium); the Agentschap voor Innovatie door Wetenschap en Technologie (IWT-Belgium); the F.R.S.-FNRS and FWO (Belgium) under the ``Excellence of Science - EOS" - be.h project n. 30820817; the Ministry of Education, Youth and Sports (MEYS) of the Czech Republic; the Lend\"ulet ("Momentum") Program and the J\'anos Bolyai Research Scholarship of the Hungarian Academy of Sciences, the New National Excellence Program \'UNKP, the NKFIA research grants 123842, 123959, 124845, 124850 and 125105 (Hungary); the Council of Science and Industrial Research, India; the HOMING PLUS program of the Foundation for Polish Science, cofinanced from European Union, Regional Development Fund, the Mobility Plus program of the Ministry of Science and Higher Education, the National Science Center (Poland), contracts Harmonia 2014/14/M/ST2/00428, Opus 2014/13/B/ST2/02543, 2014/15/B/ST2/03998, and 2015/19/B/ST2/02861, Sonata-bis 2012/07/E/ST2/01406; the National Priorities Research Program by Qatar National Research Fund; the Programa Estatal de Fomento de la Investigaci{\'o}n Cient{\'i}fica y T{\'e}cnica de Excelencia Mar\'{\i}a de Maeztu, grant MDM-2015-0509 and the Programa Severo Ochoa del Principado de Asturias; the Thalis and Aristeia programs cofinanced by EU-ESF and the Greek NSRF; the Rachadapisek Sompot Fund for Postdoctoral Fellowship, Chulalongkorn University and the Chulalongkorn Academic into Its 2nd Century Project Advancement Project (Thailand); the Welch Foundation, contract C-1845; and the Weston Havens Foundation (USA).
\end{acknowledgments}

\clearpage

\bibliography{auto_generated}

\cleardoublepage \appendix\section{The CMS Collaboration \label{app:collab}}\begin{sloppypar}\hyphenpenalty=5000\widowpenalty=500\clubpenalty=5000\vskip\cmsinstskip
\textbf{Yerevan~Physics~Institute, Yerevan, Armenia}\\*[0pt]
A.M.~Sirunyan, A.~Tumasyan
\vskip\cmsinstskip
\textbf{Institut~f\"{u}r~Hochenergiephysik, Wien, Austria}\\*[0pt]
W.~Adam, F.~Ambrogi, E.~Asilar, T.~Bergauer, J.~Brandstetter, E.~Brondolin, M.~Dragicevic, J.~Er\"{o}, A.~Escalante~Del~Valle, M.~Flechl, M.~Friedl, R.~Fr\"{u}hwirth\cmsAuthorMark{1}, V.M.~Ghete, J.~Grossmann, J.~Hrubec, M.~Jeitler\cmsAuthorMark{1}, A.~K\"{o}nig, N.~Krammer, I.~Kr\"{a}tschmer, D.~Liko, T.~Madlener, I.~Mikulec, E.~Pree, N.~Rad, H.~Rohringer, J.~Schieck\cmsAuthorMark{1}, R.~Sch\"{o}fbeck, M.~Spanring, D.~Spitzbart, A.~Taurok, W.~Waltenberger, J.~Wittmann, C.-E.~Wulz\cmsAuthorMark{1}, M.~Zarucki
% ...
\vskip\cmsinstskip
% ...
1:~Also at~Vienna~University~of~Technology, Vienna, Austria\\
% ...
\end{sloppypar}
\end{document}